\documentclass[%
reprint,
amsmath,amssymb,
aps,
longbibliography,
]{revtex4-2}

\usepackage[dvipsnames]{xcolor}
\usepackage{dcolumn}
\usepackage{bm}
\usepackage{romannum}
\usepackage[utf8]{inputenc}
\usepackage{overpic}
\usepackage{xr-hyper}
\usepackage{hyperref}
\usepackage{placeins}
\usepackage{amsmath,amsfonts,amssymb,bbold,siunitx,textgreek}

\hypersetup{
    colorlinks,
    linkcolor={blue!80!black},
    citecolor={blue!50!black},
    filecolor={red!50!black},
    urlcolor={blue!80!black}
}

\newcommand{\highlight}[1]{%
 \par\noindent
 \colorbox{white}{%
 \parbox{1.5em}{%
 #1
 }%
}}
\newcommand{\figref}[2][]{Figure~\ref{#2}\textcolor{blue!80!black}{#1}}
\newcommand{\sfigref}[2][]{Figure~\ref{#2}\textcolor{blue!80!black}{#1}}
\newcommand{\sref}[2][]{\ref{#2}\textcolor{blue!80!black}{#1}}
\renewcommand{\eqref}[1]{Eq.~(\ref{#1})}
\newcommand{\secref}[1]{Section~\ref{#1}}
\newcommand{\suppref}[1]{Appendix \ref{#1}}
\renewcommand{\vec}[1]{\mathbf{#1}}
\renewcommand{\matrix}[1]{\underline{\underline{\mathbf{#1}}}}
\newcommand{\eq}{\,{=}\,}
\newcommand{\deq}{\,{:=}\,}
\newcommand{\eqd}{\,{=:}\,}

\newcommand{\identity}[1]{\mathbb{1}_{#1}}

\begin{document}
\title{Unconventional bound states in the continuum from metamaterial induced electron-acoustic plasma waves}

\author{Wenhui~Wang}
\email{wenhui.wang@unifr.ch}
\author{Antonio~G\"{u}nzler}
\author{Bodo~D.~Wilts}
\author{Matthias~Saba}
\email{matthias.saba@unifr.ch}
\homepage{\\https://www.ami.swiss/physics/en/groups/plasmonic-networx/}
\affiliation{ Adolphe Merkle Institute, University of Fribourg, Chemin des Verdiers 4, 1700~Fribourg, Switzerland.}

\begin{abstract}\noindent
Photonic bound states in the continuum are spatially localised modes with infinitely long lifetimes that exist within a radiation continuum at discrete energy levels.
These states have been explored in various systems where their emergence is either guaranteed by crystallographic symmetries or due to topological protection.
Their appearance at desired energy levels is, however, usually accompanied by non-BIC resonances, from which they cannot be disentangled.
Here, we propose a new generic mechanism to realize bound states in the continuum that exist by first principles free of other resonances and are robust upon parameter tuning.
The mechanism is based on the fundamental band in double-net metamaterials, which provides vanishing homogenized electromagnetic fields.
We predict two new types of bound states in the continuum:
\textit{i}) generic modes confined to the metamaterial bulk, mimicking electronic acoustic waves in a hydrodynamic double plasma, and \textit{ii}) topological surface bound states in the continuum.
\end{abstract}

\maketitle


\section{\label{sec:intro}Introduction}

As early as 1929, von Neumann and Wigner constructed spatially localized bound states in the continuum (BICs) as solutions of the single particle Schr\"odinger equation with energies above the associated potential \cite{1929Uber,PhysRevA.11.446}.
BICs are, however, general wave phenomenona and are therefore not restricted to quantum mechanics and have recently been demonstrated in a variety of classical wave systems \cite{Hsu2016,marinica2008bound,hsu2013observation,molina2012surface,chen2019singularities,zhen2014topological,doeleman2018experimental,guo2017topologically,yang2014analytical,lyapina2015bound,yu2020high}. Among these different demonstrations, photonic BICs based on light-matter interactions have recently gained particular attention and have been exploited for applications in vortex beam generation  \cite{huang2020ultrafast,wang2020generating}, high-$Q$ resonators and lasing \cite{kodigala2017lasing,koshelev2018asymmetric,PhysRevLett.119.243901,ha2018directional,doeleman2018experimental}.

To date, photonic BICs involve two mechanisms: (i) symmetry guaranteed mode mismatching and (ii) destructive interference of a topological origin \cite{Hsu2016}. 
These BICs, while fundamentally interesting, are of limited use in practice as they are readily destroyed by small perturbations \cite{Liu2019,PhysRevLett.125.053902} and coupling to higher diffraction orders \cite{sadrieva2017transition}.
Furthermore, a large number of non-BIC modes with a high density of states exists spectrally close to these BICs \cite{marinica2008bound},
which considerably hinder the broadband exploitation of BICs.

\begin{figure*}
    \includegraphics[width=\textwidth]{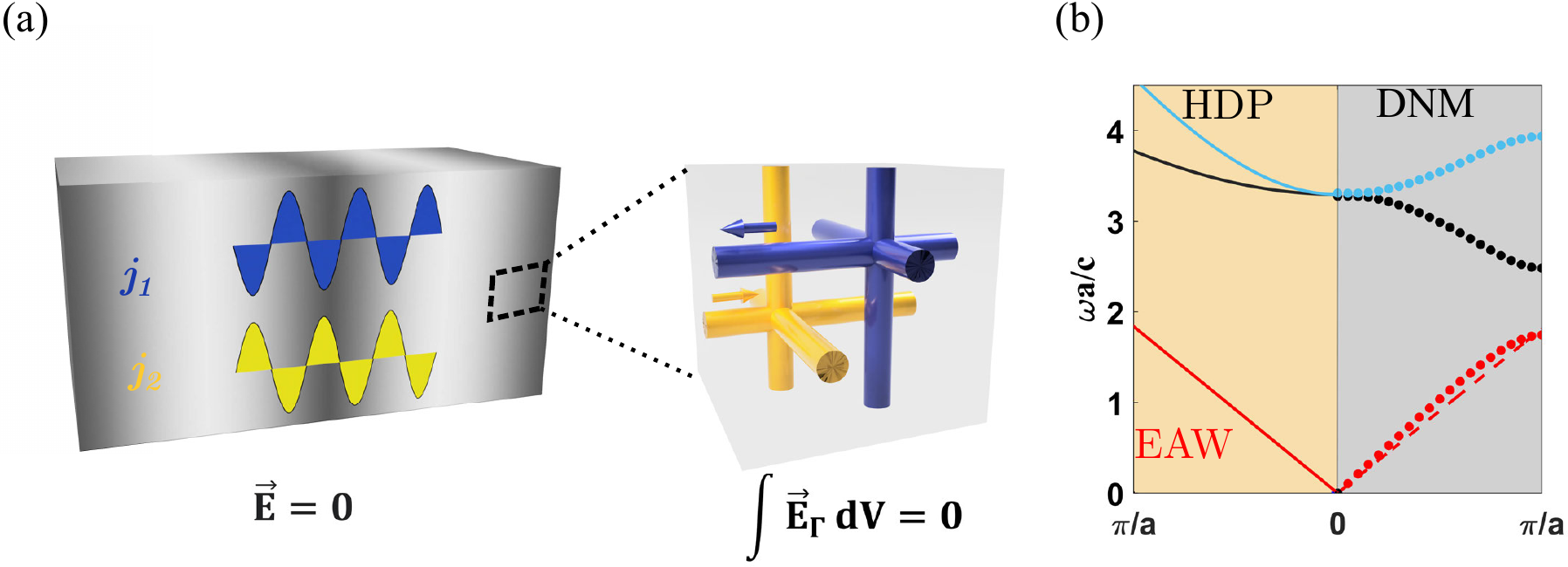}
    \caption{(a) Schematic diagram of an EAW mode in a HDP fluid. The blue and yellow counter-propagating sinusoids arise from two separate charge carriers, resulting in a vanishing field. The DNM consists of two interwoven percolating metallic nets, supporting the two counter-propagating currents. In the electrostatic limit, the 
    DNM is equivalent to an HDP, resulting in 
     a homogenized electromagnetic field that vanishes at the $\Gamma$ point. (b) Dispersion relation of an HDP with $\kappa\eq \frac{1}{\sqrt{3}}$ on the left,  featuring a light-like electromagnetic wave (cyan), a Langmuir wave (black), and an electron-acoustic wave (red). The corresponding simulated band structure of an unperturbed DNM is shown as dots on the right. The three  modes resemble those of the HDP. The slope of the fundamental band is in excellent agreement with the EAW band of HDP model (red dashed line).}
    \label{fig:1}
\end{figure*}

Here, we propose a generic hydrodynamic double fluid plasma (HDP) effective medium approach to generate new types of photonic BICs in double-net metamaterials (DNMs) consisting of two intertwining, spatially separated metallic networks \cite{Chen2018,powell2021dark}.
While the discussion in this paper is restricted to Cartesian wire meshes as shown \figref{fig:1}, we show that the HDP model equally applies to other DNM geometries in \suppref{sec:srs-dia} (\sfigref{fig:srs} and \sfigref{fig:dia}).
In contrast to existing designs, the proposed BICs are formed by pure charge fluctuations of the electron acoustic wave (EAW)-like modes in DNMs, which are strikingly robust against a large class of perturbations. They exist within a broad spectral range free of non-BIC resonances and their frequency spacing can be freely tuned by the thickness of the structure, reaching spectrally dense BICs for an optically thick DNM slab.
Additionally, new Zak-phase protected topological suface bound states in the continuum (TSBICs) are discovered in DNMs.
Both bulk BICs and TSBICs exist at the $\Gamma$-point of the in-plane Brillouin zone within the light-cone for DNM slabs, yet they are fully decoupled from radiation.

\section{Electronic acoustic wave in double-net metamaterials  \label{sec:bulk}}

The new types of BICs originate from a pure charge wave, which does not produce electromagnetic fields. Although impossible in conventional optical materials, this property can be found in non-Maxwellian
materials \cite{shin2007three}, for example in an HDP.
The hydrodynamic motion of two charge carriers in combination with Maxwell's equations describes the fluid-like dynamics of an effective  double plasma in an electromagnetic field. 
The full HDP model is derived in \suppref{sec:HDP}. In short, we restrict the discussion to a non-dissipative model with only two free parameters: the plasma frequency $\omega_{\mathrm{p}i}\deq \sqrt{n_{0i}q_i^2/m_i}$ and the thermal pressure parameter $\kappa_{i}\deq \sqrt{\gamma p_i/(n_{0i} m_ic^2)}$ for the two charge carriers labelled $i\eq1,2$. These depend on the equilibrium charge carrier density $n_{0i}$, the charges $q_i$, the effective masses $m_i$, the equilibrium pressure $p_i$, the adiabatic exponent $\gamma$ and the speed of light in vacuum $c$.

Through a plane-wave ansatz and by linearizing the HDP model \cite{wang2020photonic,gao2016photonic}, the generally complex problem is transformed into a $14$-dimensional linear eigenproblem for the eigenfrequency $\omega$. As shown in \figref[(b)]{fig:1}, from high to low frequencies, a two-fold degenerate electromagnetic transverse band and a longitudinal Langmuir band are obtained, both of which are hyperbolic starting at $\omega\eq(\omega_{\mathrm{p}1}^2+\omega_{\mathrm{p}2}^2)^{1/2}$ for wave number $k\eq0$, and a linear electron acoustic wave (EAW) band emanating from $\omega\eq0$.
The slope of the EAW band is $\pm\frac{c}{\sqrt{2}}\sqrt{\kappa_{1}^2+\kappa_{2}^2}$, as analytically shown by $k{\cdot}p$ theory \cite{dresselhaus2008group} (\suppref{sec:effective}).
Regardless of $\omega_{\mathrm{p}1,2}$, when $\kappa_{1}\eq{\kappa_{2}}$,
the EAW consists of two charge waves drifting in opposite direction, with current densities
$\vec{j}_1\eq{-}\vec{j}_2\parallel\vec{k}$ and vanishing electromagnetic fields $\vec{E}\eq\vec{H}\eq0$ (\figref[(a)]{fig:1}). Due to its zero electromagnetic nature and the resulting decoupling from the electromagnetic vacuum modes, the EAW naturally forms BICs.

To circumvent the difficulties in creating a natural double-plasma \cite{Gary1985,Montgomery2001,Hellberg2000}, we instead employ a finite DNM lattice, whose fully connected wire morphology effectively provides a freely moving double-plasma with finite equilibrium charge carrier density and Coulomb pressure, leading to a behavior that resembles a plasma with an exact electronic thermal pressure.  Note that in DNMs, $\kappa_{1,2}$ are largely determined by the connectivity of the wires, which leads to $\kappa_{1,2}\,{\approx}\,\frac{1}{\sqrt{3}}$ (\suppref{sec:pcu}).  
The DNM structures discussed in this paper are composed of two single nets of fully connected metal cylinders along the cubic axes with radii $r_{1,2}$, which may be different for the two nets.
We use the notation $P\deq(r_1/a,r_2/a,R)$ to represent the network parameters, normalized by the lattice constant $a$.
$R$ is the offset of the two networks that are shifted by $\vec{S}\deq (1,1,1)^\intercal\, Ra$.
Two parameter sets are discussed in this manuscript,
\begin{align*}
    P_1 := \left(\frac{2}{25},\frac{2}{25},\frac{1}{3}\right) &\text{ and}\quad P_2(\psi) := \left(\frac{2}{25},\frac{2}{25\psi},\frac{1}{2}\right) \text{ .}
\end{align*}

\begin{figure}
    \includegraphics[width=\columnwidth]{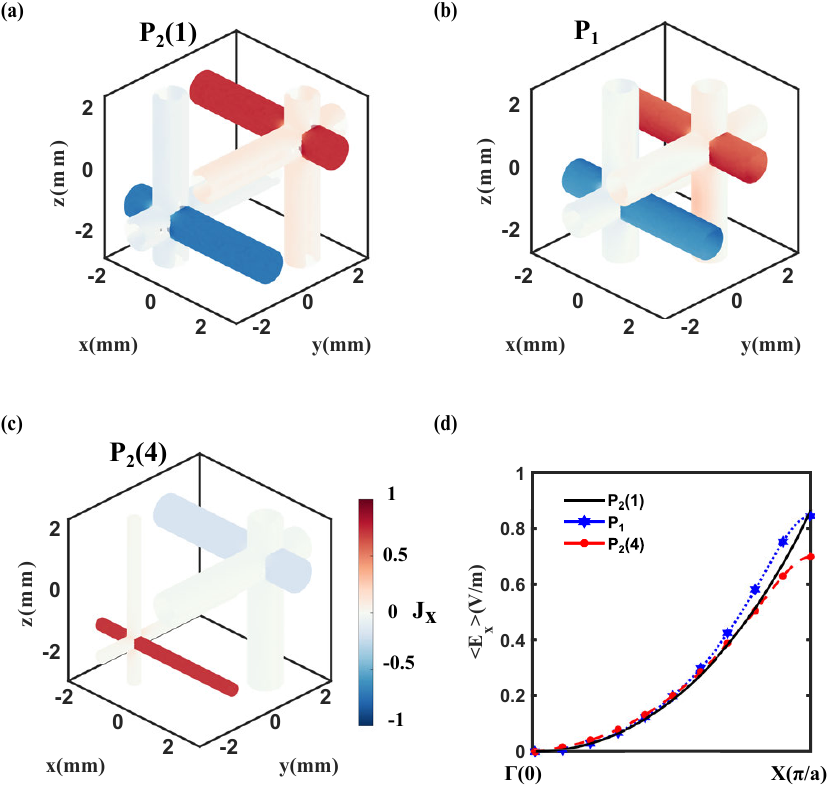}
    \caption{ (a)-(c) Surface current density component $J_{x}$, normalized by its maximum value in the unit cell, for three DNMs at a small wave vector of $k_{x}\eq 0.1 \frac{\pi}{2a}$, $k_y\eq k_z\eq0$.
    (d) Absolute values of the homogenized electric field along the $\Gamma{-}X$ ($x$ direction) for the longitudinal EAW band, for the three parameter sets in (a)-(c), $P_2(1)$ (black), $P_1$ (blue), and $P_2(4)$ (red).} 
    \label{fig:2}
\end{figure}
 
The numerically simulated band structure of the $P_1$ DNM 
is illustrated in \figref[(b)]{fig:1}, featuring a linearly dispersed fundamental band (red dotted line) that emanates from the $\Gamma$-point at zero frequency.
This result not only qualitatively matches the EAW band in the HDP model, but the dispersion is indeed in good agreement with the analytical prediction from \suppref{sec:effective} $\omega\eq\frac{1}{\sqrt{3}} ck$ (red dashed line).
In the HDP model, the slope of the EAW band is largely independent of the plasma frequencies $\omega_{\mathrm{p}1,2}$ (\suppref{sec:HDP}), which applies to a variety of DNM realizations (\suppref{sec:params}), as confirmed by numerical simulations  (\sfigref{fig:s4}, \sref{fig:s5}).
Analogously to the EAW in the HDP model, the two networks in the DNM carry opposing currents, indicated by arrows in \figref[(a,b)]{fig:1}. 
Simulated surface current densities $J$ are shown in \figref[(a,b)]{fig:2} for DNMs with equivalent radii and different offsets ($P_1$ and $P_2(1)$).
They indeed exhibit almost exactly opposite $J_x$ in the two networks at $k_x\eq \pi/a$ and $k_{y,z}\eq 0$.
For different radii ($P_2(4)$,  \figref[(c)]{fig:2}), a weaker current density is observed on the larger radius network, whose integrated magnitude equals the opposing current density of the lower-radius network.
This implies that the total current remains zero, irrespective of the network radii, consistent with the HDP model.

\begin{figure}
    \includegraphics[width=\columnwidth]{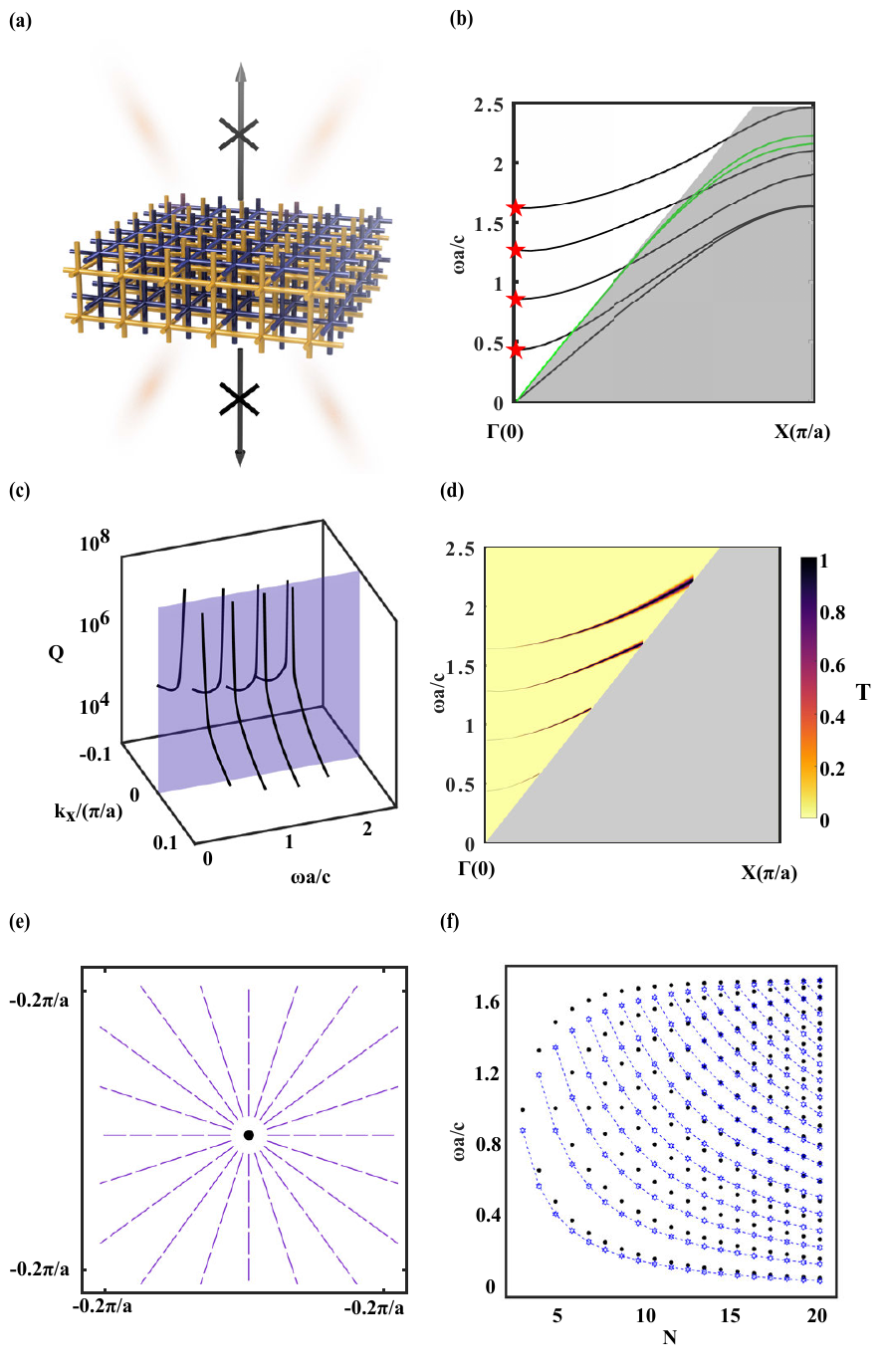}
    \caption{
    (a) Sketch of a DNM slab embedded in vacuum.
    (Quasi-) BICs on (emerging from) the $\Gamma$-point do not (weakly) radiate energy.
    (b) Quasi-normal mode bandstructure of an $N\eq5$ unit-cell thick $P_1$ DNM slab.
    The black quasi-BIC bands emanate from the red BIC points at $\Gamma$ and consist of two EAW bulk modes.
    The two green bands correspond to spoof-plasmon surface modes.
    The region outside of the light cone is shaded in grey.
    (c) The $Q$-factors of the four quasi-BIC modes are large at finite $\vec{k}$ and diverge at $\Gamma$.
    (d) Transmittance of an incoming plane wave within the light cone.
    The transmission is $0$, except close to the quasi-BIC bands, on which it is $1$.
    (e) Linear polarization direction of the far-field in $\vec{k}$-space.
    (f) Frequency of the BIC modes for $N$ unit cell thick slabs.
    The simulated frequencies (black dots) match the analytical prediction from the HDP model (blue) well.
    }
    \label{fig:3}
\end{figure}

To investigate the nature of the microscopic electromagnetic fields of the EAW-like modes in DNMs, we compute the unit-cell-averaged electric field components of the fundamental mode, normalized by the averaged electric field intensity.
The averaged transverse components vanish over the entire Brillouin zone (\sfigref{fig:s3}), while the longitudinal component quadratically increases with the wave vector when moving away from $\Gamma$ (\figref[(d)]{fig:2}).
The fundamental modes in DNMs are therefore  quasi-longitudinal with $\vec{E}\parallel\vec{k}$.
We conclude that the HDP model is a valid effective medium description for DNMs near the static limit and that DNMs provide a metamaterial realization of HDPs that require extremely stringent restrictions in natural plasmas, impeding their laboratory observation.
We finally note that due to the non-Maxwellian  nature of the EAW that lacks electromagnetic fields, the introduction of an effective permittivity seems misleading. 
In this context, for the $P_2(1)$ network, a previously reported homogenization approach \cite{Lannebere2020} leading to an effective permittivity $\varepsilon_\mathrm{h}$ for the DNM only accounts for the Langmuir mode residing at high frequencies. 
In contrast to standard metamaterial homogenization approaches, the HDP model developed here yields a fully consistent effective medium description of the fundamental DNM band and provides an accurate prediction of both the phase velocity and the associated fields.



\section{Bound states in the continuum in double net metamaterials \label{sec:BIC}}

Here, we show that the EAW character of the bulk modes in DNMs naturally leads to the formation of a new type of BIC, if it is embedded in vacuum in a slab configuration, as illustrated in \figref[(a)]{fig:3}.
In the HDP picture, a standing wave solution exists at certain frequencies where the EAW solutions can be superposed so that the normal component of $\vec{j}_i$ vanishes at the slab boundaries.
As the EAW fields satisfy $\vec{E}\parallel\vec{k}$, they perfectly decouple from the radiative vacuum modes \cite{joannopoulos2011photonic}.
To illustrate the above prediction, we simulate a $5$-layer slab of a $P_1$ DNM (\sfigref{fig:5layergeoms}).
We apply Bloch-periodic boundary conditions in the $x{-}y$-plane and scattering boundary conditions in the $z$-direction, and solve for the complex-valued frequencies.
As predicted, multiple BICs with infinite lifetimes are found, located at the $\Gamma$-point, marked by red pentagrams in \figref[(b)]{fig:3}.
The frequencies of the BICs are well approximated by the standing wave solution of the EAW with a predicted phase velocity of $\frac{1}{\sqrt{3}}\,c$ and hard wall boundary conditions ($\vec{j}_i\eq0$).
The EAW band is well below the first Wood anomaly, so that the higher Bragg orders in the vacuum Rayleigh basis are evanescent and non-radiative \cite{cryst5010014,rayleigh1907dynamical}.
Because the EAW bulk band is spectrally isolated from the other non-evanescent bulk bands at low frequencies, the BICs are free of non-BIC slab modes in the entire spectral range of the EAW band.
In contrast, BICs in photonic crystals are usually mixed with non-BIC modes \cite{Hsu2016}.

At finite wave vectors away from the $\Gamma$-point, quasi-BICs are formed since the EAW fields are not strictly orthogonal to the vacuum plane waves, which results in a weak coupling to the vacuum states.
These quasi-BICs thus have a very large, but finite, quality factor. The corresponding $Q$-factor values of the multiple quasi-BIC solutions in the vicinity of the $\Gamma$-point are shown in \figref[(c)]{fig:3}. The physics is reminiscent of a Fabry-P\'erot resonator with mirrors that are perfectly reflecting at normal incidence and otherwise weakly transparent. In scattering experiments, this system yields a simple transmittance spectrum with a Lorentzian line shape at finite wave vectors close to the BIC frequency. The centre frequency forms a band that emanates from the BICs at the $\Gamma$-point and hyperbolically blue-shifts with increasing wave vector, while the width of the Lorentzian gradually broadens. The simulated transmittance spectrum along the $\Gamma$-$X$ direction is shown in \figref[(d)]{fig:3}.
The polarization of the quasi-BIC far-field is oriented in the direction of the wave vector, forming a `star' pattern in 2D momentum space, as shown in \figref[(e)]{fig:3}.
At the $\Gamma$-point where the BICs reside, this generates a singular point where the polarization cannot be defined since the far-field vanishes.
Around this point, the polarization pattern evidently exhibits a non-trivial topological winding number \cite{zhen2014topological}.
In our case, this topological vortex structure of the polarization field is a direct consequence of the quasi-longitudinal nature of the EAW modes and is reminiscent of the electric field lines created by a point charge.

In photonic crystal slabs and metasurfaces, BICs are essentially symmetry protected and can be destroyed by infinitesimal symmetry-breaking perturbations \cite{Liu2019,PhysRevLett.125.053902}.
Strikingly, the new BICs formed in DNMs are immune to global symmetry breaking and are only prone to perturbations in the individual networks.
The DNM slab with $P_1$ has indeed only one point symmetry, the $(1\overline{1}0)$ mirror symmetry, which is insufficient to create symmetry-protected BICs in photonic crystal slabs that require at least two mirror planes.  
The stability of the DNM-based BICs can be attributed to the surface current densities $\vec{J}$ on the metallic wires.
For the EAW mode at $\vec{k}\eq \frac{\pi}{a} (0.05,0,0)^\intercal$, the currents $J_{x}$ on the two metallic cylinders along the $x$-direction are exactly opposite (\figref[(a--c)]{fig:2}). $J_{y}$ and $J_{z}$ on each single net resemble electrical quadrupoles, whose form is determined exclusively by the the single net geometry, and which are found irrespective of the global symmetry (\suppref{sec:SI4}, \sfigref{fig:s6}).
Since the interaction between the currents on the two nets is negligibly small in the static limit, the total radiated field from the quadrupole-like current vanishes, irrespective of their relative positions, radii, \textit{etc}. 
This ultimately leads to BICs that are robust against global symmetry breaking. 
This robustness of the BIC modes is demonstrated in \suppref{sec:SI5} through simulations of DNM slabs with various parameters (\sfigref{fig:s7}).
As shown in \suppref{sec:SI6}, the BICs can, however, be lifted by breaking the local $C_{4v}$ symmetry of individual nets (\sfigref{fig:s8}), spawning circularly polarized points in the far-field (\sfigref{fig:s9}).


Finally, since the BICs originate from the standing wave solutions of the EAWs in the DNM slabs, the number of BIC modes and their frequency spacing are exclusively determined by the thickness of the DNM slab.
A full theoretical model that predicts both the quasi-BIC bands including their quality factors (\sfigref{fig:analytical_QNM}) and the transmission spectrum (\sfigref{fig:analytical_T}) is presented in \suppref{sec:analytical}.
In particular, the model yields BIC modes that are equidistantly spaced at frequencies of $\frac{\omega a}{c}\eq l\frac{\kappa \pi}{N}$, where $Na$ is the thickness of the DNM slab and $N\,{>}\, l\,{\in}\,\mathbb{N}$.
Extrapolating the linear dispersion relation of the HDP model to the Brillouin zone boundary, we expect $N{-}1$ BIC modes over the EAW band, which perfectly matches the full wave simulation results of the $P_1$ DNM in \figref[(f)]{fig:3}.
Generally, the theoretical results are in good agreement with the simulations even though the equidistant frequency spacing is of course perturbed in real DNMs.

\section{\label{sec:topoBS}Topological surface bound states in the continuum}

\begin{figure}
    \includegraphics[width=\columnwidth]{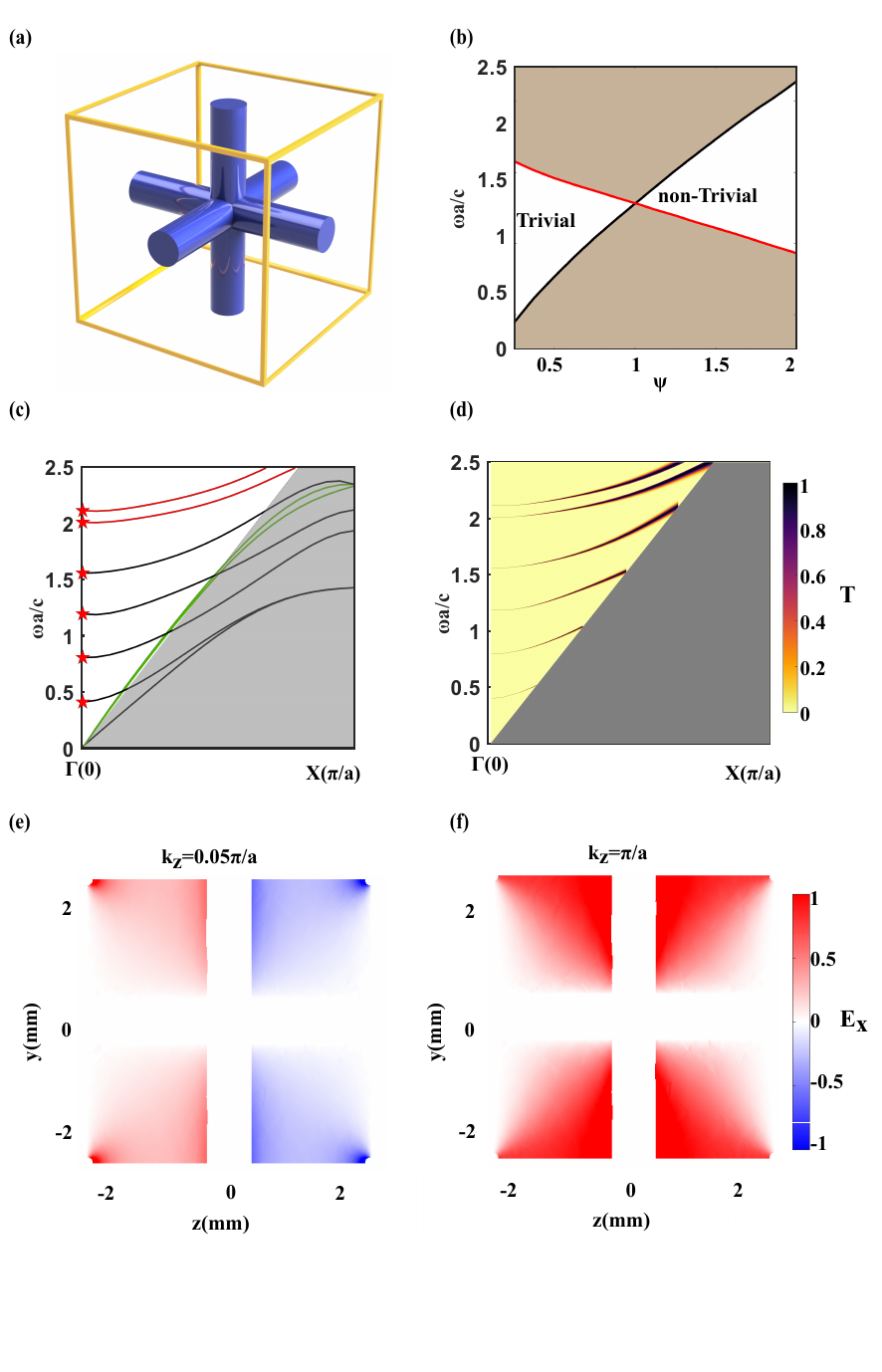}\vspace{-3em}
    \caption{
    (a) Geometry of a $P_2(4)$ DNM consisting of a thick (blue) and a thin (yellow) wire network.
    (b) Projected surface bands at $\Gamma$ as a function of $\psi\eq r_1/r_2$.
    A topological phase transition occurs for $\psi\eq 1$, where the bandgap closes. 
   (c) Quasi-normal slab modes as in \figref[(b)]{fig:3}.
   Additional to the black quasi-BICs and the green surface bands, two TSBIC bands are found inside the projected bulk gap (red).
    (d) Transmittance of an incoming plane wave within the light cone as in \figref[(c)]{fig:3}.
    The quasi-TSBICs give rise to high-$Q$ transmission similar to the bulk BIC bands.
    (e,f) $P_2(4)$ EAW electric field component $E_{x}$ across the $(100)$ plane in the center of the unit cell for $k_x\eq k_y\eq0$, and (e) $k_{z}\eq 0.05 \frac{\pi}{a}$ and (f) $k_{z}\eq \frac{\pi}{a}$. The parity change, even in (e) and odd in (f), yields a Zak phase of $\pi$ for the EAW band.
    }
    \label{fig:4}
\end{figure}

Our theory builds on the HDP homogenization model that, while a powerful tool to understand the bulk modes and the formation of BICs, cannot capture the discrete periodicity of  DNMs.
This discrete lattice structure, however, bears additional features.
For example, as predicted in \cite{Chen2018}, a primitive lattice translation of the $P_2(1)$ 
network with BCC symmetry interchanges the two individual networks and results in an EAW band emanating from the corner of the Brillouin zone at the $H$-point ($\vec{k}\eq(2\pi/a,0,0)$), as shown in \sfigref{fig:s10} \footnote{When perturbed from $\psi\eq1$, the symmetry is lowered to the simple cubic $Pm\overline{3}m$ space group. The EAW band thus emanates from $\Gamma$ and it seems as if the mode discontinuously jumps from left-handed to right-handed propagation. The conundrum is resolved by realizing that the main Bragg amplitudes of the Bloch wave remain at non-zero reciprocal lattice vectors, causing a left-handed behavior that is not immediately evident from the bandstructure.}.
A lattice symmetry with an underlying highly symmetric isogonal point group can often be associated with non-trivial topological properties \cite{po2017symmetry,tang2019comprehensive,chen2012symmetry,saba2017group}.
As shown below, the lattice symmetry of the DNMs gives rise to a non-trivial Zak phase and topological surface bound states in the continuum (TSBICs).
TSBICs exist at the interface between the DNM slab and vacuum within a radiation continuum, in contrast to typical topological surface states, which reside in bulk bandgaps of two neighbouring domains \cite{Lu2016}.

The emergence of TSBICs can be understood by examining the family of $P_2(\psi)$ DNMs.
A schematic diagram of a $P_2(4)$ DNM is shown in \figref[(a)]{fig:4}, with the corresponding bandstructure in \sfigref{fig:s11}.
For $\psi\eq1$ (i.e.\ equal network radii), the structure has the full body centered cubic (BCC) symmetry.
As a result, the EAW and the Langmuir bands meet at $X$, since they stem from one band that is back-folded half-way between the $\Gamma$ and the $H$-point of the BCC Brillouin zone \cite{SETYAWAN2010299}.
For $\psi\,{\ne}\, 1$, a bandgap opens as predicted by the geometrical perturbation theory  described in \cite{saba2017group}.
In the presence of inversion and time reversal symmetry, the band topology can be characterized by a $\mathbb{Z}_2$ (binary) topological index, called the Zak phase \cite{zak1989berry}, which is best known for the quantized bulk polarization of electrons in solids.
The Zak phase is the well established Berry phase \cite{doi:10.1098/rspa.1984.0023} evaluated across a 1D Brillouin zone, which is topologically a circle and therefore closed.
It is generally a continuous number on $[0,2\pi)$ that depends on the choice of the unit cell.
The Zak phase is, however, $\mathbb{Z}_2$-quantized, with values of $0$ and $\pi$, if the unit cell has a parity symmetry in its center \cite{PhysRevX.4.021017}.
Therefore, an offset of $R\eq\frac{1}{2}$ between the two nets is necessary to preserve the Cartesian mirror planes in the $O_h$ point symmetry of $Im\overline{3}m$, allowing a Zak phase classification.

In our case, the Zak phase is obtained by examining parity of the electromagnetic Bloch fields at the high symmetry points of the 1D Brillouin zone \cite{van2016topological}.
We make use of the $(001)$ mirror plane (operator $\varsigma_z$) in the centre of unit cell in \figref[(a)]{fig:4}.
The
fields must have even ($1$) or odd (${-}1$) parity with regard to $\varsigma_z$ at the $\Gamma$ and $X$-points, where the wave vector is invariant under $\varsigma_z$.
The Zak phase of the band has been shown to be the complex argument of the product of both parities \cite{van2016topological}.
A topological phase transition from trivial to non-trivial Zak states is observed in the projected band structure shown in \figref[(b)]{fig:4} at $\psi\eq1$, where the network has the higher BCC symmetry and the bandgap closes at the $X$-point of the simple cubic Brillouin zone.
The field parities for $\psi\eq4$ are illustrated by the $z$-components of the periodic part of the electric Bloch field of the EAW mode close to $\Gamma$ \footnote{On $\Gamma$ itself, the numerical full wave solution yields a multi-dimensional eigenspace, including spurious modes, with no simple 1D even/odd representation of the mirror group.} in \figref[(e)]{fig:4} and at $X$ in \figref[(f)]{fig:4}.
For a DNM slab terminated at its thin wires, we therefore predict, and indeed see, topological surface states within the bandgap, shown as red lines in \figref[(c)]{fig:4}. Note that the DNM slab is truncated slightly away from the mirror plane through the thin wires as shown in \sfigref{fig:5layergeoms}(b), to avoid slicing the thin wire.
As discussed in \suppref{sec:SI8}, an energy difference between the two surface states is observed, which is caused by hybridization of the surface states on either side of the slab through its finite thickness (\sfigref{fig:TSBICJz}).
The interaction is weakened by increasing the slab thickness, as shown in \sfigref{fig:s11}(b).

The topological surface bands in \figref[(c)]{fig:4} are evidently associated with high $Q$-factor quasi-TSBIC bands revealed by the reflectivity plot in \figref[(d)]{fig:4}.
TSBICs are, in stark contrast to conventional photonic topological surface states \cite{Lu2016}, confined through the EAW mechanism and do not require a surrounding topologically trivial band-gap material.
When different terminations are applied to the top and bottom surfaces, so that only one surface obtains TSBICs, no resonance is found in the reflection/transmission amplitudes, since no hybridization, which provides a direct transmission channel, occurs.
The presence of a single quasi-TSBIC band manifests itself, however, in a resonance in the reflection phase (\sfigref{fig:Rphase}).
Despite earlier reports of surface BICs in terms of the intrinsic continuum in the material \cite{molina2012surface}, our TSBICs constitute the first example of BICs of topological origin which are decoupled from free space radiation. Controlling the coupling between the topological surface states with free space light could pave the way towards both fundamental scattering properties of topological states and their application, for example in  topological antenna arrays \cite{lumer2020topological}.

\section{\label{sec:conclusions}Conclusions}
We have demonstrated a new deterministic route towards photonic bound states in the continuum in double-net metamaterials.
A hydrodynamic double fluid plasma serves as an effective medium model for these materials.
The hydrodynamic double plasma yields an analytical solution with a new type of quasi-longitudinal EAW band, which gives rise to photonic bound states in the continuum that are robust even against strong geometrical perturbations and symmetry breakings.

Moreover, topological phase transitions are found in double-net metamaterials, accompanied by topological surface bound states in the continuum, protected by a Zak phase topological index. The marriage between the bound-state behavior and the non-trivial topological phase enables the lossless existence of TSBICs even for open boundaries.
Our findings endow using double-net metamaterials as a new platform for the realization of photonic bound states in the continuum by employing a new topological strategy.
Advances in 3D fabrication, particularly the 3D printing of metals, enables the manufacture of the proposed double-net metamaterials.
While an experimental verification of our predictions is thus readily available at microwave frequencies,  we intend to extend our findings to higher frequencies, allowing to, for example, facilitate quantum emission from the coherent spontaneous emission regime, over photonic Bose-Einstein condensation to lasing applications.

\begin{acknowledgements}
    We would like to thank Ullrich Steiner for fruitful discussions and general support. We acknowledge funding by the Swiss National Science Foundation through the Spark Grant 190467 and the Project Grant 188647.
\end{acknowledgements}

\renewcommand{\thefigure}{A\arabic{figure}}
\setcounter{figure}{0}
\appendix

\section{\label{sec:srs-dia}Different double network geometries}
While our theory above has been rigorously derived for the pcu double network only, the two main physical features are inherent to the plasmonic double network topology and symmetry rather than the particular geometrical realization.
These features are the slope of the EAW band in the low frequency limit and the longitudinal nature of the homogenized electric field.
We here demonstrate that both features are indeed be observed for two other established double net geometries: the balanced double gyroid (srs-c, $Ia\overline{3}d$ symmetry) and double diamond (dia-c, $Pn\overline{3}m$ symmetry) morphologies (abbreviations as in \cite{rcsr}).

Let us first re-examine the longitudinal nature from a symmetry perspective.
In the quasi-static limit, a non-trivial solution to the Poisson equation requires the two nets to be on different potentials \cite{Chen2018}.
Three situations can generally be distinguished:
\begin{enumerate}
    \setlength{\itemsep}{0pt}
    \item the two nets are only interchanged by one or more space group elements $S$ with point symmetry part that is not the identity\footnote{In other words, we have $S\eq(P|\vec{t})$ in Seitz notation with a point symmetry $P{\ne}\identity{}$ and a translation part $\vec{t}$, which might be non-zero (screw axis or glide mirror for example).}\label{point},
    \item the two nets are interchanged by a primitive lattice translation\footnote{$S\eq(\identity{}|\vec{T})$ in Seitz notation, where $\vec{T}$ must be a primitive lattice vector.}\label{trans},
    \item the two nets are no symmetry copy of one another. This case is always possible for unbalanced double nets as for example the qtz-qzd double net \cite{markande2018chiral}.
\end{enumerate}
The cubic examples discussed here all belong to the symmetric cases \ref{point} and \ref{trans}.
In these cases, the square $S^2$ of the symmetry operation that exchanges nets evidently maps the two nets onto themselves.
The potential is therefore unchanged by $S^2$ as the individual nets are short-cut in the quasi-static limit.
As a consequence, $S$ yields a multiplication of the potential by ${-}1$ and the two nets must hence be on opposite potential.
In case \ref{point}, the Bloch character is trivial and the EAW band emanates at zero frequency at the $\Gamma$-point.
In case \ref{trans}, the Bloch character is ${-}1$ with respect to a primitive lattice translation and the EAW hence emanates at zero frequency at the corner of the Brillouin zone.

\begin{figure}
    \begin{overpic}[width=\columnwidth]{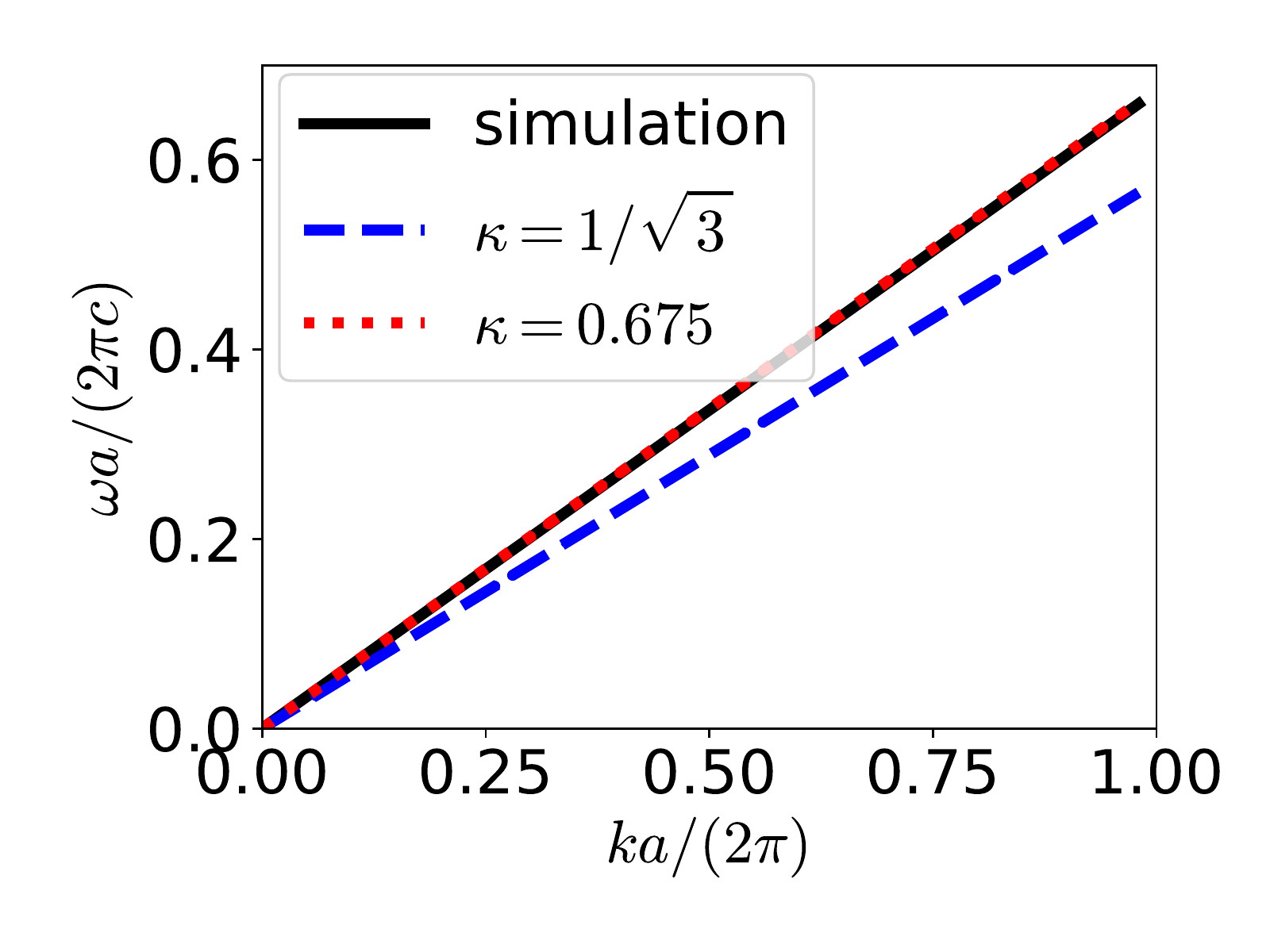}
        \put(1,70){(a)}
    \end{overpic}
    \begin{overpic}[width=\columnwidth]{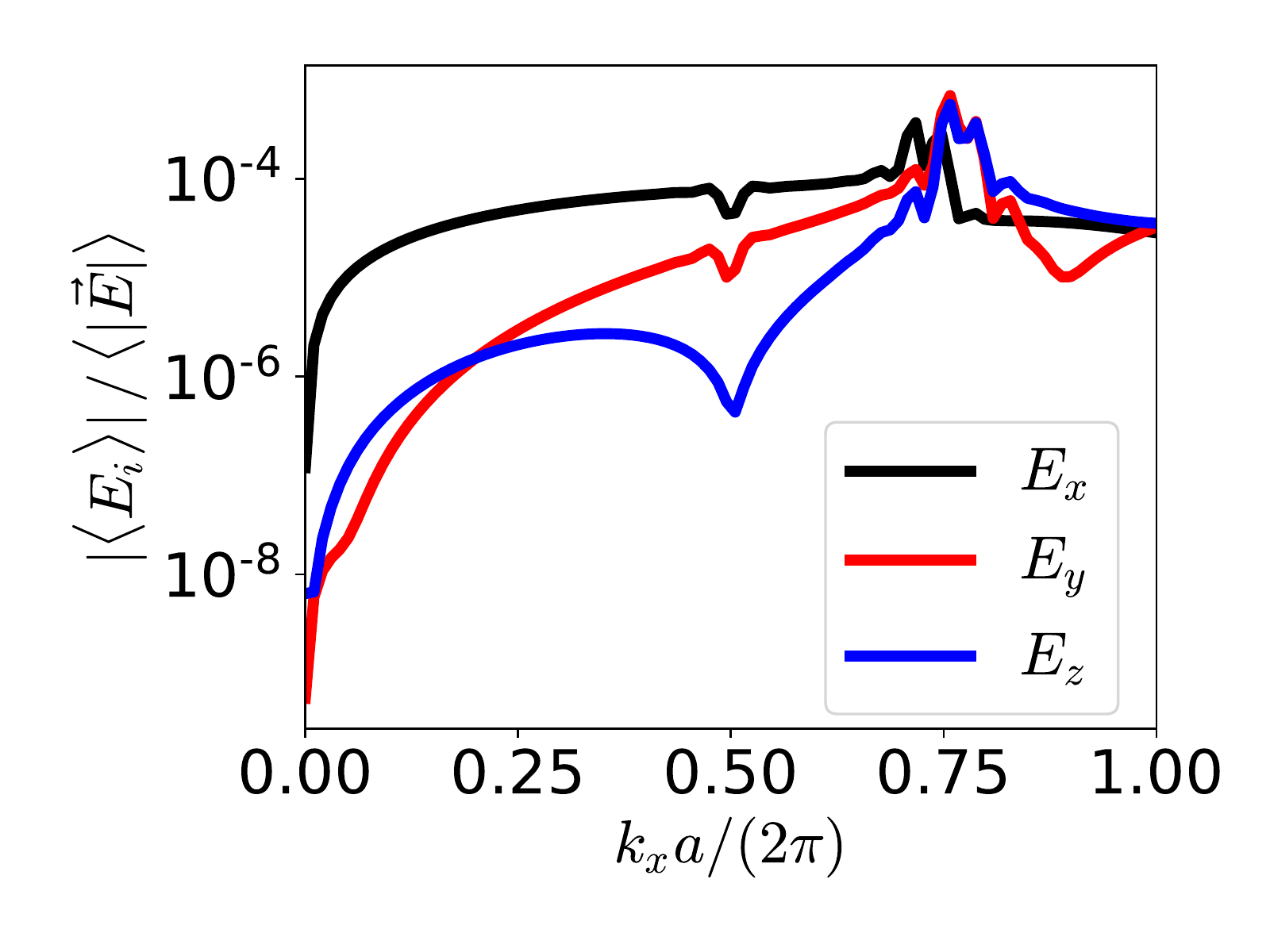}
        \put(1,70){(b)}
    \end{overpic}
    \caption{
    EAW mode characterization for the srs-c double net along the $\Gamma$-$H$ path (parametrized by $\vec{k}\eq k_x\,\vec{e}_x$).
    (a) Dispersion relation compared to the HDP prediction with limiting ($1/\sqrt{3}$) and radius corrected $\kappa\eq0.675$.
    (b) Averaged field components normalized by averaged field intensity parallel to $\vec{k}$ ($[100]$) and along perpendicular high symmetry directions.
    }
    \label{fig:srs}
\end{figure}

With these preliminary considerations in mind, we first revisit the pcu-c double net.
As the pcu-c evidently belongs to case \ref{trans} above, with a primitive body-centered cubic lattice translation interchanging nets, the EAW band emanates at zero frequency at the $H$ point \cite{SETYAWAN2010299}.
Since $H$ lies along the cubic $\langle100\rangle$ direction, any state at the center of the surface Brillouin zone (with vanishing lateral $\vec{k}$) of a $(100)$ inclinated slab is therefore made by a superposition of counter-propagating EAW modes along $\Gamma$-$H$.
The behavior in the quasi-static limit now further yields the irreducible representation, or \emph{irrep}, of the electric field with respect to the group of the wave vector \cite{dresselhaus2008group}, which is the $C_{4v}$ point group in this case.
Since all elements of $C_{4v}$ do not interchange networks and the individual networks are on constant potential in the quasi-static limit, the EAW irrep is trivial.
Therefore, the homogenized field cannot have a component perpendicular to the wave vector direction, as these would transform with a non-trivial 2D $E$ irrep.
The fields must be longitudinal from a symmetry perspective over the whole EAW and Langmuir bands, which are glued together and form one band.

\begin{figure}
    \begin{overpic}[width=\columnwidth]{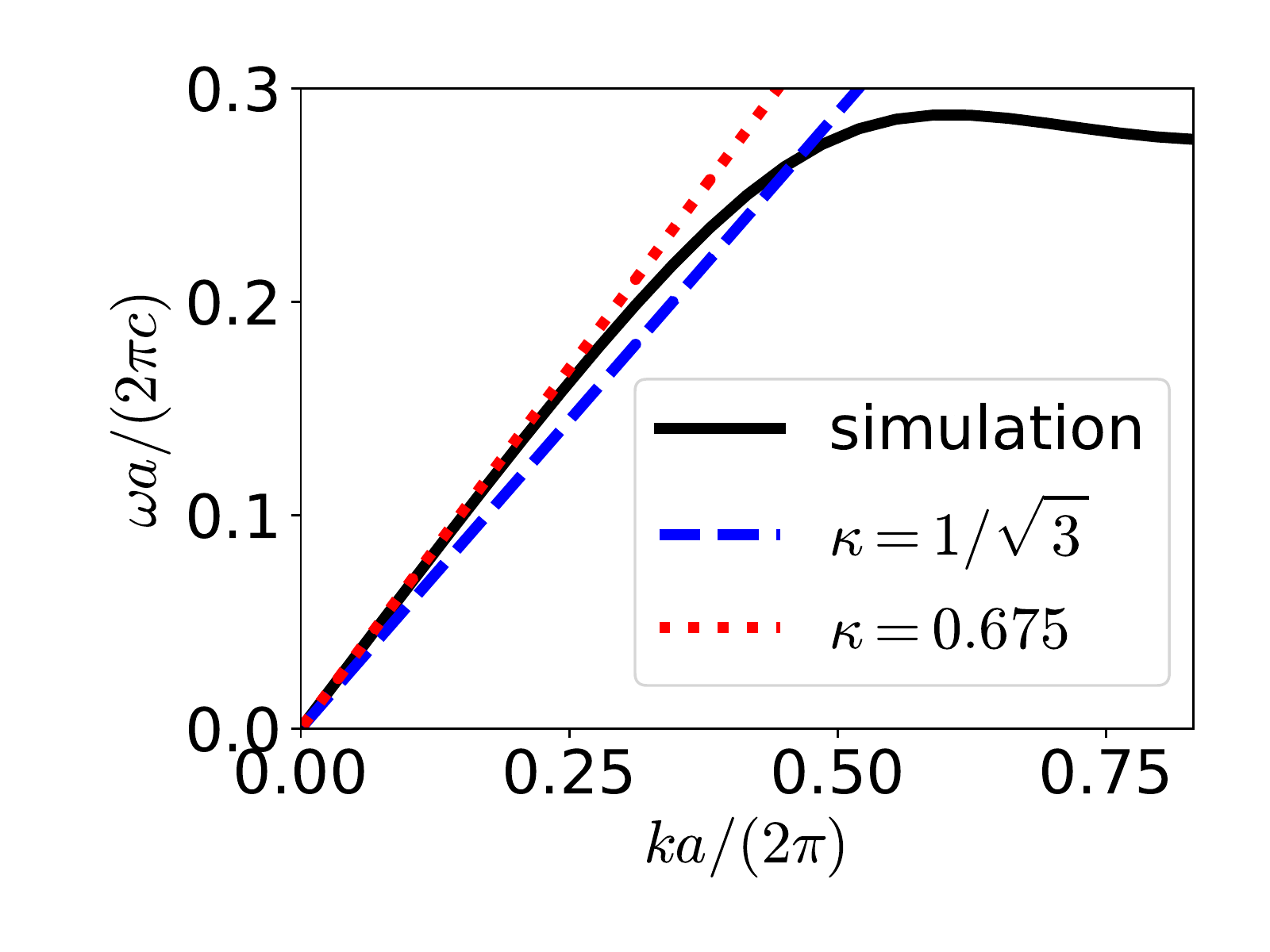}
        \put(1,70){(a)}
    \end{overpic}
    \begin{overpic}[width=\columnwidth]{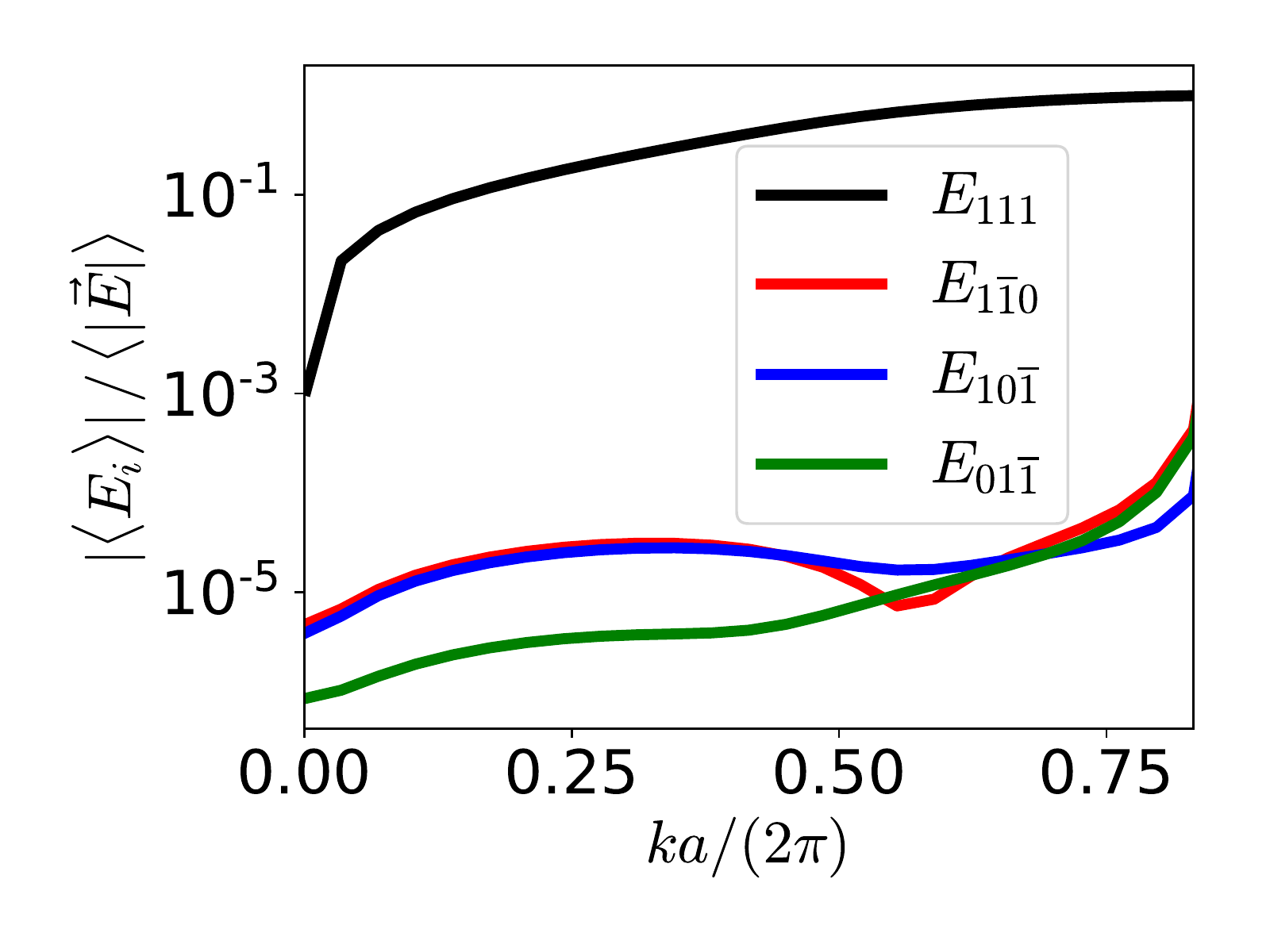}
        \put(1,70){(b)}
    \end{overpic}
    \caption{
    EAW mode characterization for the dia-c double net along the $R$-$\Gamma$ path (parametrized by $\vec{k}\eq(\frac{\sqrt{3}\pi}{a}{-}k)\frac{1}{\sqrt{3}}(1,1,1)^\intercal$). (a) Dispersion relation compared to the HDP prediction with limiting ($1/\sqrt{3}$) and radius corrected $\kappa\eq0.675$.
    (b) Averaged field components normalized by averaged field intensity parallel to $\vec{k}$ ($[111]$) and along perpendicular high symmetry directions.
    }
    \label{fig:dia}
\end{figure}

The srs-c is of case \ref{point} above, and the EAW band emanates at the $\Gamma$-point of the body-centered cubic Brillouin zone at zero frequency.
The bandstructure along $\Gamma{-}H$ ($\vec{k}\eq k\,\vec{e}_x$ with $k\in[0,2\pi/a]$) for wire radius of $0.02a$ is shown in \figref{fig:srs} (a).
It agrees perfectly with the HDP prediction for the corrected pressure parameter of $\kappa{\approx}0.675$ for the wire radius of $0.02a$ over the whole band.
This can be understood as follows:
The group of the wave vector is again $C_{4v}$, which is isomorphic to the abstract group $G_8^4$ in \cite{bradley1972mathematical}.
Considering the quasi-static potentials, the EAW band must be even with respect to the $C_42$ screw rotation and the $C_2$ rotation (which map each single net onto itself), and odd with respect to the glide mirror planes (which interchange networks).
It therefore has the 1D representation $R_2$ in $G_8^4$.
At the $H$ point, the EAW joins a $6$-fold degenerate point, which transforms according to the $R_{14}$ representation in $G_{96}^4$, which splits equally into the $4$ 1D and one 2D representations along $\Gamma$-$H$: $R_{14}(G_{96}^4)\eq\sum_{i\eq1}^5 R_i(G_8^4)$.
This allows the $R_2$ representation to form a pair with its time reversal symmetric $R_4$ representation to form a back-folding at $H$ with finite and opposite slope of the two bands.

As in the pcu-c scenario, the longitudinal EAW behavior is protected by symmetry.
Since the group of the wave vector contains a $C_{42}$ screw axis and diagonal glide mirrors that both contain a shift along the propagation direction, only the non-symmorphic wallpaper subgroup $p2gg$ (number $8$ in \cite{brock2016international}), with isogonal $C_{2v}$ point group, can be employed to make a general prediction regarding the homogenized fields.
Considering the quasi-static potentials again, we obtain an even symmetry classification with respect to the $C_2$ symmetry in $p2gg$.
This classification evidently does not permit homogeneous field components perpendicular to the direction of the wave vector.
The odd symmetry behavior with respect to the glide mirror symmetry in $p2gg$ on the other hand prohibits a homogenized field component along $\vec{k}$.
For the srs-c, we therefore expect a truly vanishing homogenized electric field that extends over the whole EAW band.
This vanishing field is demonstrated in \figref{fig:srs} (b) within the numerical precision of the simulations (note the logarithmic $y$-axis).

Similar to the pcu-c, the dia-c network with simple cubic $Pn\overline{3}m$ symmetry belongs to case \ref{trans} above.
The EAW band therefore emanates at the simple cubic $R$-point \cite{SETYAWAN2010299}.
Consequently, the EAW band lies outside of the light cone along the cubic $\langle100\rangle$ and $\langle110\rangle$ directions and cannot lead to BICs.
For a slab with $(111)$ inclination, however, the EAW band along the path connecting $R$ with $\Gamma$ ($\vec{k}\eq (\frac{\sqrt{3}\pi}{a}{-}k)\frac{1}{\sqrt{3}}(1,1,1)^\intercal$ with $k\in[0,\frac{\sqrt{3}\pi}{a}]$) lies at the center of the surface Brillouin zone.
The low frequency dispersion agrees again perfectly with the HDP prediction with corrected $\kappa{\approx}0.675$ as shown in \figref{fig:dia} (a).
It therefore yields BICs by the same mechanism as exploited in the pcu-c case with slab inclination of $(100)$, assuming a longitudinal homogenized electric field.
Such a field is guaranteed by the trivial symmetry classification of the mode with respect to the $C_{3v}$ group of the wave vector\footnote{Even though $Pn\overline{3}m$ is non-symmoprhic, the elements of the group of the wave vector (labelled 1, 5, 9, 38, 43 and 48 in \cite{brock2016international}) are pure mirrors and rotations through the crystallographic origin.}, which only allows a homogenized field parallel to $\vec{k}$.
This longitudinal nature is once again demonstrated through full wave simulations in \figref{fig:dia} (b).

\section{\label{sec:HDP}Plasma and pcu network modes}
\subsection{Single plasma fluid model \label{sec:single_plasma}}
We begin with a single plasma fluid model for a warm plasma \cite{fitzpatrick2015plasma}. This model consists of a set of three equations describing the charge carrier (CC) dynamics,
\begin{subequations}
\begin{align}
    \frac{\partial n}{\partial t}+\nabla \cdot (n\vec{u})&=0\label{eq:chargeconv}\\
    m\left(\frac{\partial\, n\vec{u}}{\partial t}+\nabla\cdot [n\vec{u}\otimes\vec{u}]\right) &=
    -\nabla P+\rho\vec{E} \label{eq:NavierStokes}\\
    \frac{\mathrm{d}}{\mathrm{d}t}\left(Pn^{-\gamma}\right)&=0 \label{eq:energyconv}\text{ .}
\end{align}
\label{eq:CCequations}
\end{subequations}
\eqref{eq:chargeconv} is the continuity equation of CC conservation, where $n(\vec{r},t)$ is the CC volume density field and $\vec{u}(\vec{r},t)$ is the CC velocity field.
\eqref{eq:NavierStokes} is the Navier-Stokes equation for the electrically excited CC liquid, where we ignore shear forces, the non-linear magnetic field contribution in the Lorentz force density, and gravitational contributions. In this equation, $\otimes$ is the outer product, $m$ the effective mass of the charge carriers, $P(\vec{r},t)$ the CC liquid pressure field, and $\rho(\vec{r},t)\deq q n(\vec{r},t)$ the CC charge density.
We note that a kinetic pressure term is present in any liquid. Here, $P$ should be understood as the total mesoscopic pressure based on the microscopic statistical ensemble of CCs, including inter-particle Coulomb interactions.
Finally, \eqref{eq:energyconv} is a local adiabatic equation with the adiabatic index $\gamma$, which can be rigorously derived from an energy balance equation in the limit where the flow velocity is much larger than vicious and thermal diffusion velocities \cite{fitzpatrick2015plasma}.
These fluid equations are closed with the Maxwell curl equations, which describe the electro-magnetic field dynamics. In Lorentz-Heaviside units,
\begin{subequations}
    \begin{align}
        \nabla\times\vec{E} &= -\frac{1}{c}\frac{\partial \vec{H}}{\partial t} \label{eq:Faraday}\\
        \nabla\times\vec{H} &= \frac{1}{c} \left( \frac{\partial\vec{E}}{\partial t} + \vec{j}     \right)\label{eq:Ampere}\text{ ,}
    \end{align}
    \label{eq:Maxwell}
\end{subequations}
with $\vec{E}(\vec{r},t)$ and $\vec{H}(\vec{r},t)$ the electric and magnetic fields, respectively, $c$ the speed of light in vacuum, and $\vec{j}(\vec{r},t)\deq\rho(\vec{r},t)\vec{u}(\vec{r},t)$ the current density of the plasma charge carriers.

We now linearize \eqref{eq:CCequations} assuming small variations in the density and pressure
\begin{align*}
    n(\vec{r},t) &:= n_0 + \delta n(\vec{r},t) \\
    P(\vec{r},t) &:= P_0 + \delta P(\vec{r},t)
\end{align*}
and thus small velocities $\vec{u}(\vec{r},t)$ to arrive at
\begin{subequations}
\begin{align}
    \frac{\partial \delta n}{\partial t}+n_0 \nabla \cdot \vec{u}&=0\\
    m\, n_0 \frac{\partial\, \vec{u}}{\partial t} &=
    -\nabla \delta P+\rho\vec{E} \\
    \frac{\mathrm{d}}{\mathrm{d}t}\left(\delta P - \gamma \frac{P_0}{n_0}\delta n \right)&=0 \label{eq:energyconv_linear} \text{ .}
\end{align}
\label{eq:CCequations_linear}
\end{subequations}
Since $\delta P\eq0$ if $\delta n\eq0$, \eqref{eq:energyconv_linear} is solved by $\delta P\eq \gamma P_0\,\delta n/n_0$. Finally, since \eqref{eq:Maxwell} and \eqref{eq:CCequations_linear} are now both spatio-temporally homogeneous and linear in all fields, we can make a plane-wave ansatz $n(\vec{r},t)\deq n \exp\{\imath(\vec{k}\cdot\vec{r}{-}\omega t)\}$, \textit{etc.}, such that the field symbols represent their constant coefficients from this point on, to obtain
\begin{subequations}
    \begin{align}
        c\,\vec{k}\times\vec{E} &= \omega \vec{H} \label{eq:Faraday_pw}\\
        -c\,\vec{k}\times\vec{H} &= \omega \vec{E} + \imath\vec{j} \label{eq:Ampere_pw} \\
    \omega \delta n &=  n_0 \vec{k} \cdot \vec{u} \label{eq:chargeconv_pw}\\
     \omega m n_0 \vec{u} &=
    -\vec{k}\, \frac{\gamma P_0}{n_0}\delta n+\imath q n_0 \vec{E} \label{eq:NavierStokes_pw}\text{ .}
    \end{align}
    \label{eq:pwequations}
\end{subequations}
We now introduce the plasma frequency $\omega_\mathrm{p}^2 \deq n_0 q^2/m$, the plasma wave number $k_\mathrm{p}\deq\omega_\mathrm{p}/c$, and the dimensionless pressure parameter $\kappa^2 \deq \gamma P_0 / (n_0 m c^2)$, to obtain a $\vec{k}$-family of linear Hermitian eigenproblems
\begin{equation}
    \mathcal{H}(\vec{k})\,\vec{v} = \frac{\omega}{\omega_\mathrm{p}}\,\vec{v}
    \label{eq:eigenproblem}
\end{equation}
for the eigenvalue $\omega/\omega_\mathrm{p}$. The eigenvector is defined as
\begin{equation*}
    \vec{v} := 
    \begin{pmatrix}
    \vec{E} \\ \vec{H} \\ \frac{\kappa}{k_\mathrm{p}}q\, \delta n \\ \frac{1}{\omega_\mathrm{p}}\vec{j}
    \end{pmatrix}\text{ .}
\end{equation*}

The dimensionless matrix operator
\begin{equation}
    \mathcal{H} :=
    \begin{pmatrix}
        \mathcal{H}_{\text{field}} & \mathcal{H}_{\text{int}} \\
        \mathcal{H}_{\text{int}}^\dagger & \mathcal{H}_{\text{fluid}}
    \end{pmatrix} \text{ ,}
    \label{eq:Hamiltonian1}
\end{equation}
which we from now on refer to as Hamiltonian, has the sub-blocks
\begin{align*}
    \mathcal{H}_{\text{field}} := \frac{1}{\imath k_\mathrm{p}}
        \sigma_y\otimes \matrix{K}_\times
    \text{ , }
    &\mathcal{H}_{\text{int}} := 
    \begin{pmatrix}
        1 \\ 0
    \end{pmatrix}
    \otimes
    \begin{pmatrix}
        \vec{0} & -\imath\identity{3}
    \end{pmatrix}
    \text{ ,}\\
    \text{and}\quad&\mathcal{H}_{\text{fluid}} := \frac{\kappa}{k_\mathrm{p}}
    \begin{pmatrix}
        0 & \vec{k}^\dagger \\ \vec{k} & \matrix{0}
    \end{pmatrix}\text{ .}
\end{align*}

Here, we have used the Pauli spin matrix $\sigma_y$, the vector product matrix $\matrix{K}_{\times} \vec{w}\deq \vec{k}{\times}\vec{w}$, the Kronecker (tensor) product $\otimes$, the zero vector $\vec{0}$ and matrix $\matrix{0}$ of dimension $3$, and the identity matrix $\identity{d}$ of dimension $d$; $\matrix{A}^\dagger$ denotes the Hermitian adjoint of $\matrix{A}$.

\begin{figure}
    \includegraphics[width=\columnwidth]{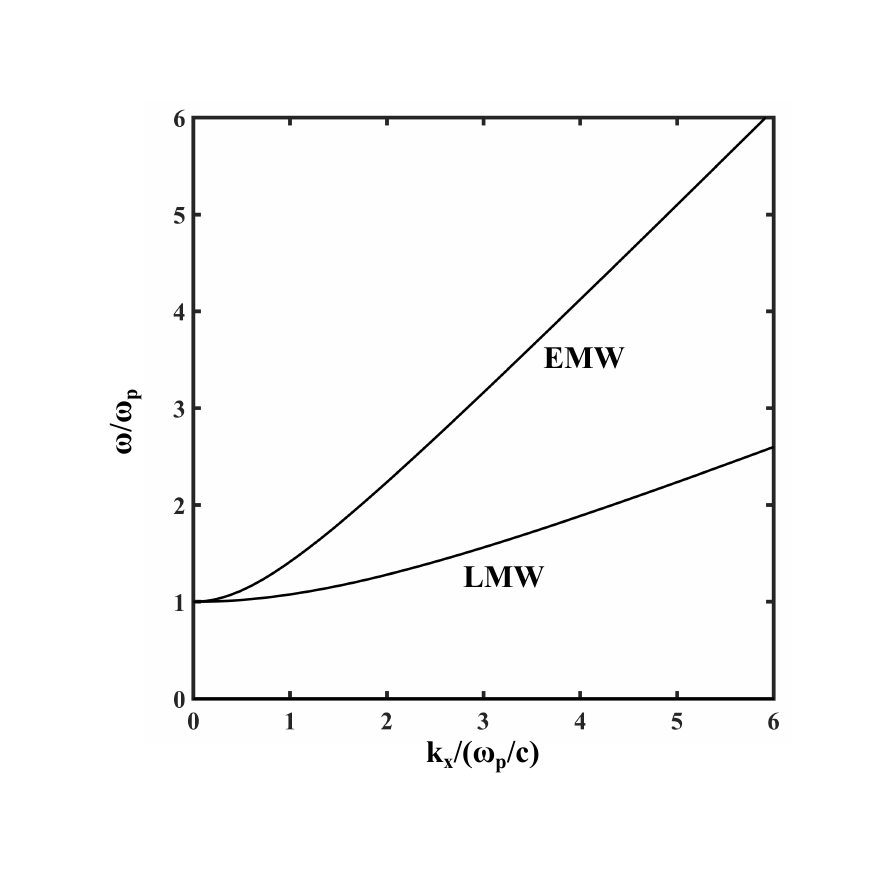}
    \caption{Dispersion relation of a single plasma fluid with thermal coefficient $\kappa\eq0.4$. It has a longitudinal Langmuir (LMW) and a doubly-degenerate electromagnetic wave (EMW) band.}
    \label{fig:s1}
\end{figure}

The problem is of course isotropic, so that we can choose the coordinate frame such that $\vec{k}\eq k\,\vec{e}_z$ without loss of generality. The eigenproblem \eqref{eq:eigenproblem} predicts a four-fold degenerate zero-frequency band $\omega(k)\eq 0$. The associated eigenspace is spanned by two longitudinal spurious modes, which violate the Maxwell divergence equations, and the static magnetic field solutions generated by a transverse current density
$$
\begin{pmatrix}
    H_x \\ H_y
\end{pmatrix} = 
-\frac{k_\mathrm{p}}{k} \sigma_y
\begin{pmatrix}
    j_x \\ j_y
\end{pmatrix}
$$
at constant charge $\delta_n\eq0$ and vanishing electric field.

Note that the system is particle-hole symmetric \cite{jin2016topological}, that is, for every eigenvector $\vec{v}$  of $H\vec(k)$ with energy $\omega$, there is one particle-hole symmetric eigenvector  ${\vec{v}}^*$ with energy $-\omega$. 
Thus we just show the positive frequency bands in \figref{fig:s1}. They follow the longitudinal mode Langmuir dispersion\footnote{Note that we do not normalize fields and choose the electric and magnetic field without dimension (instead of their standard cgs unit $\text{g}^{1/2}\text{cm}^{-1/2}\text{s}^{-1}$) here for convenience.}
\begin{equation}
    \frac{\omega_l}{\omega_p} = \sqrt{1+\left(\frac{\kappa k_l}{k_\mathrm{p}}\right)^2}
    \label{eq:long_disp}
\end{equation}
with fields $\vec{E}\eq\vec{e}_z $, $\vec{H}\eq0$, $  \delta n \eq \frac{\imath k_l}{q}$, and 
$\vec{j} \eq \imath \omega_l \vec{e}_z$; 
and the two-fold degenerate transverse light mode dispersion
\begin{equation}
    \frac{\omega_t}{\omega_p} = \sqrt{1+\left(\frac{k_t}{k_\mathrm{p}}\right)^2}
    \label{eq:trans_disp}
\end{equation}
with fields $\vec{E}_\pm \eq \vec{e}_x{\pm}\imath \vec{e}_y$, $\vec{H}_\pm \eq \frac{ck_t}{\omega_t}\left(\vec{e}_y{\mp}\imath\vec{e}_x\right)$, $\delta n \eq 0$, and $\vec{j} \eq \frac{2\imath\omega_p^2}{\omega_t}\vec{E}_\pm$ \text{.}

\subsection{The metal pcu wire-mesh \label{sec:pcu}}

At low frequencies in the microwave regime, most metals act like prefect electrical conductors (PECs), which do not support propagating modes in the bulk.
Employing plasmonic metal wires, we can however reduce the CC density to introduce an effective plasma freqeuncy for excitations in the wire direction, while maintaining the lossless PEC character \cite{https://doi.org/10.1002/mop.10512}.
A 3D metallic pcu \cite{rcsr} wire-mesh with sub-wavelength period therefore effectively generates a lossless plasma at microwave excitations. 
Based on \cite{Maslovski2009}, we here formally show that the electro-dynamic theory of the metallic pcu network in the long wavelength limit, is indeed equivalent to the single plasma model with a constant pressure parameter $\kappa=1/\sqrt{3}$.


\begin{figure*}
    \includegraphics[width=0.8\textwidth]{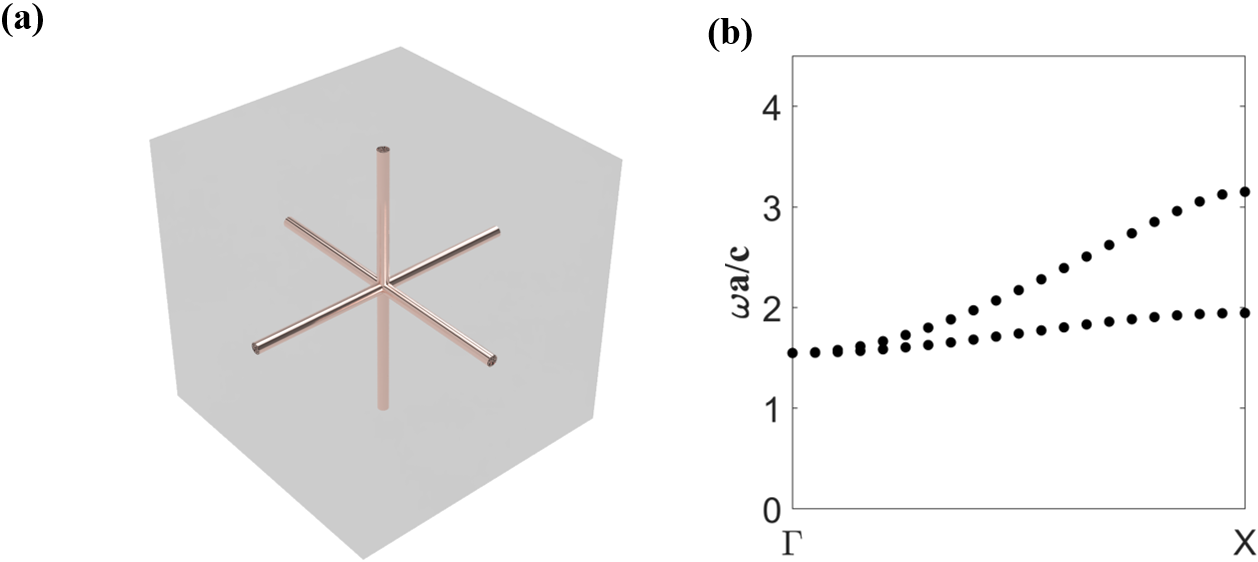}
    \caption{(a) Single metallic wire-mesh with radius $r\eq 0.02 a $. (b) Dispersion relation of (a).} 
    \label{fig:s2}
\end{figure*}

We use the macroscopic constitutive relations from \cite{Maslovski2009} that have been derived for a square lattice of aligned wires (along $\vec{e}_z$) in the homogenization picture.
We start with the current $I(z)$ and the electric potential $V(z)$ between neighbouring wires.
A comparison of the magnetic flux through a rectangle of infinitesimal height with the corresponding boundary integral (Amp\'ere's law in integral form) yields (assuming a plane-wave form)\footnote{Note that we here use the physics convention for the monochromatic ansatz $\exp\{-\imath\omega t\}$ and Lorentz-Heaviside units in contrast to \cite{Maslovski2009}.}
\begin{equation}
     k_z V =\omega L I -\imath\langle E_z\rangle,
     \label{eq:Ampere_macro}
\end{equation}
with the self-inductance per unit length of the wires in the mesh
$   L = \frac{1}{2\pi c^2} \log\left(\frac{a^2}{4 r(a-r)}\right)$
that is a simple consequence of the Biot-Savart law for the individual wires and only depends on the lattice constant (nearest neighbour distance) $a$ and the wire radius $r$. 
The potential between neighbouring wires is on the other hand related to the linear charge density $\lambda(z)$ via the linear relation $\lambda \eq C V$ with the effective capacitance per unit length $C\eq1/(c^2 L)$ that is obtained using Gauss' law. 
Combining this relation with charge conservation $k_z I\eq\omega\lambda$ leads to
\begin{equation}
    k_z I = \omega C V \text{.}
    \label{eq:CC_macro}
\end{equation}
\eqref{eq:Ampere_macro} and \eqref{eq:CC_macro} can be readily generalized to a 3D connected pcu wire-mesh: \eqref{eq:Ampere_macro} simply acquires full vector form with $k_z\mapsto \vec{k}$, $I\mapsto a^2\langle\vec{j}\rangle$, and $\langle E_z\rangle\mapsto\langle \vec{E}\rangle$. 
The linear charge density is effectively reduced by a factor of $1/3$ as the charges spread over three wires in each direction in the unit cell and the pcu network is fully connected. A more rigorous treatment for wires of finite radius reveals that this factor is indeed better approximated by \cite{1420782}
\begin{subequations}
\begin{align}
    \kappa^2 &= \frac{1+2\frac{k_\mathrm{p}^2}{k_1^2}}{3}\text{ ,}     \label{eq:kappa}\\
    &\text{ with the plasma wavenumber }\notag\\
    \frac{k_\mathrm{p} a}{2\pi} &= \left[\sum_{(m,n)\in\mathbb{N}^2\setminus(0,0)}
    \frac{J_0^2 \left(\frac{2\pi r}{a}\sqrt{m^2+n^2}\right)}{m^2+n^2}
    \right]^{(-1/2)} \text{,}\label{eq:kp}\\
    \text{ and }\frac{k_1 a}{2\pi} &= \left[\sum_{m\in\mathbb{N}\setminus0}
    \frac{J_0^2 \left(\frac{2\pi r}{a} m\right)}{m^2}
    \right]^{(-1/2)}\text{ .}
\end{align}
\end{subequations}
In practice, the wire-mesh pressure parameter falls in a range between $\kappa\,{\approx}\,2/3$ for a wire radius of $r\eq a/100$ and $\kappa\,{\approx}\,3/4$ for $r\eq a/5$.
Rearranging, we obtain
\begin{subequations}
    \begin{align}
        \imath\langle\vec{E}\rangle + \vec{k} V &= \omega a^2L \langle\vec{j}\rangle \\
        \kappa^2a^2\vec{k}\cdot\langle\vec{j}\rangle &= \omega C V \text{ .}
    \end{align}
\end{subequations}
Defining the plasma frequency as the cut-off frequency of the aligned wire mesh \cite{https://doi.org/10.1002/mop.10512} $k_\mathrm{p}^2 a^2 \deq C$, which is indeed a good approximation of the definition in \eqref{eq:kp}, and the field vector
$$
\vec{v} :=
\begin{pmatrix}
    \langle\vec{E}\rangle \\ \langle\vec{H}\rangle \\ \frac{\kappa}{k_\mathrm{p}}V \\ \langle\vec{j}\rangle/\omega_p
\end{pmatrix}\text{ ,}
$$
we formally arrive at the above eigenproblem \eqref{eq:eigenproblem} with a $\kappa$ that is confined in a small interval and only weakly depends on the geometrical parameters of the wire-mesh. We finally note that the scaled potential $\frac{\kappa}{k_\mathrm{p}} V$ equals the scaled CC density $\frac{\kappa}{k_\mathrm{p}}q\delta n$ fom \secref{sec:single_plasma}, if we associate the linear charge density along the PEC wires per unit cell area $\lambda/a^2$ with the charge density $\rho$ of the plasma, so that the field vector $\vec{v}$ is identical in both models. \figref{fig:s2} shows the bandstructure obtained through a full-wave simulation, which agrees well with the single plasma fluid model shown in \figref{fig:s1}. The two modes in \figref{fig:s2}(b) correspond to the longitudinal (LMW) and transverse (EMW) modes in \figref{fig:s1}. 

\subsection{Hydrodynamical double plasma fluid  model and plasmonic double pcu wire-mesh}

\begin{figure*}
    \includegraphics[width=\textwidth]{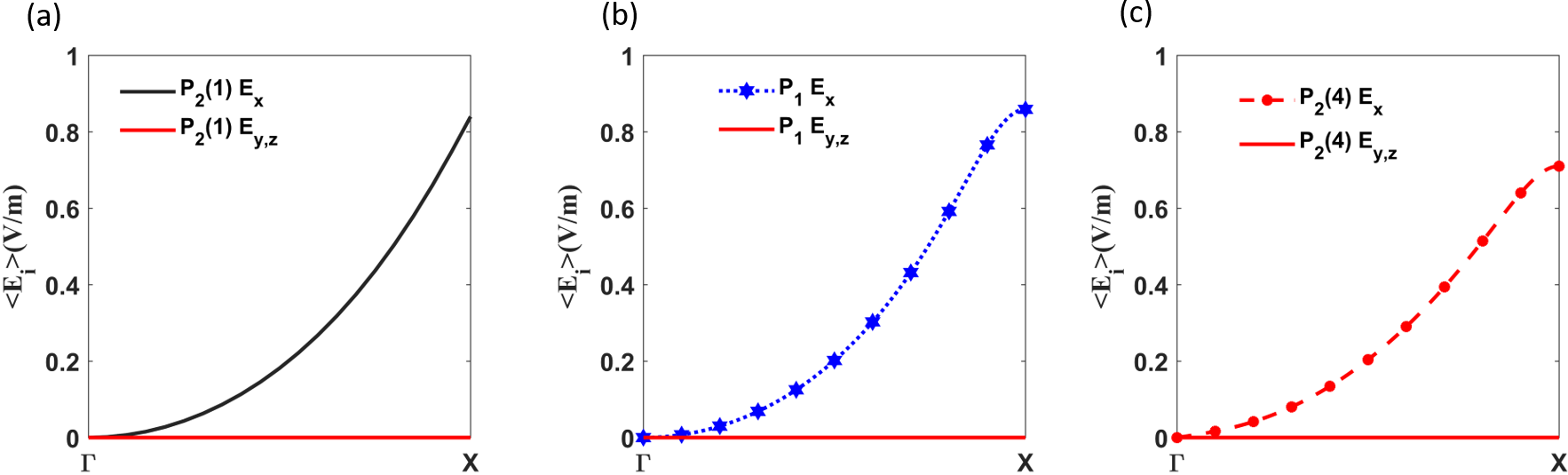}
    \caption{Absolute value of the electric field, averaged over one unit cell, for each component of the EAW bands along the high symmetry path  $\Gamma{-}X$ along the $x$ direction. The parameter choices are (a) $P_{2}(1)$;(b) $P_1$; and (c) $P_{2}(4)$. }
    \label{fig:s3}
\end{figure*}
We now consider a plasma consisting of two CC species of respective charge $q_i$, mass $m_i$, and equilibrium CC density $n_0^{(i)}$ and pressure $P_0^{(i)}$ ($i\eq 1,2$). The associated plasma frequencies, plasma wave numbers and pressure parameters are$ \omega_{\mathrm{p}i}$, $k_{\mathrm{p}i}$ and $\kappa_i$. We further assume that CC interact only amongst their own species, which is physically of course only possible in a universe where particles of a different type do not interact (via Coulomb forces or otherwise). The theory is, however, approximately valid if the interaction is weak or if the two charge carriers are separated on a mesoscopic scale.

In the prescribed situation, the corresponding fluid equations \eqref{eq:chargeconv_pw} and \eqref{eq:NavierStokes_pw} separate into two individual equations, while both current densities contribute to the Amp\`ere-Maxwell law \eqref{eq:Ampere_pw}, 
\begin{subequations}
    \begin{align}
        c\,\vec{k}\times\vec{E} &= \omega \vec{H} \label{eq:Faraday_double}\\
        -c\,\vec{k}\times\vec{H} &= \omega \vec{E}
        + \imath\sum_{i=1}^2 \vec{j_i} \label{eq:Ampere_double} \\
    \omega \delta n_i &=  n_0^{(i)} \vec{k} \cdot \vec{u}_i  
    \quad (i=1,2)\label{eq:chargeconv_double}\\
     \omega m_i n_0^{(i)} \vec{u}_i &=
    -\vec{k}\, \frac{\gamma P_0^{(i)}}{n_0^{(i)}}\delta n_i+\imath q_i n_0^{(i)} \vec{E}\quad (i=1,2) \label{eq:NavierStokes_double}\text{ .}
    \end{align}
    \label{eq:pwequations_double}
\end{subequations}

As we have shown in \secref{sec:pcu}, a PEC pcu wire-mesh acts like a single plasma in the homogenization regime, that is if the Bloch wavevector of the mode satisfies $|\vec{k}|\,{\ll}\, 2\pi/a$.
We use the parameter vector $P\deq(r_1/a,r_2/a,R/a)$ to define a pcu double network, where $r_i$ are the individual network radii and $a$ the lattice constant, and $R$ is the offset in each Cartesian direction, so that the two networks are separated by a shift vector $\vec{S}\deq (1,1,1)^\intercal\, R$.
Assuming that the networks are well separated, the DNM fields therefore obey \eqref{eq:pwequations_double} with the substitutions discussed in \secref{sec:pcu}.

The two networks generally exhibit different plasma frequencies $\omega_{\mathrm{p}i}$ that depend on $r_i$ and lie approximately between $k_\mathrm{p} a/(2\pi)\eq 1/5$ for $r\eq1/100$ and $k_\mathrm{p} \/(2\pi)\eq3/5$ for $r\eq1/5$. The pressure parameters are generally also different, but contained in a very small intervall between $\kappa\eq2/3$ and $\kappa\eq3/4$, so that it is useful to introduce $\kappa_1\deq\kappa{-}\delta\kappa$ and $\kappa_2\deq\kappa{+}\delta\kappa$ with $\delta\kappa{\ll}\kappa$.
The canonical field vector of the double wire-mesh is therefore
\begin{equation*}
    \vec{v}_{\text{DN}} :=
    \begin{pmatrix}
        \langle\vec{E}\rangle \\ \langle\vec{H}\rangle \\
        \frac{\kappa{-}\delta\kappa}{k_{\mathrm{p}1}} V_1 \\ \langle\vec{j}_1\rangle \\        
        \frac{\kappa{+}\delta\kappa}{k_{\mathrm{p}2}} V_2 \\ \langle\vec{j}_2\rangle
    \end{pmatrix}.
\end{equation*}

The eigenproblem can be written as
\begin{align}
    \mathcal{H}_\text{DN}(\vec{k})\,\vec{v}_\text{DN} = \frac{\omega}{c}\,\vec{v}_\text{DN}
    \label{eq:EVP2}
\end{align}
with the associated Hamiltonian
\begin{equation}
    \mathcal{H}_\text{DN} :=
    \begin{pmatrix}
        \mathcal{H}^\text{DN}_\text{field} & \mathcal{H}^\text{DN}_\text{int} \\
        \left(\mathcal{H}^\text{DN}_\text{int}\right)^\dagger & \mathcal{H}^\text{DN}_\text{fluid}
    \end{pmatrix} \text{ ,}
    \label{eq:Hamiltonian2}
\end{equation}
with the sub-blocks
\begin{equation*}
    \mathcal{H}^\text{DN}_\text{field} :=
        -\imath \sigma_y\otimes \matrix{K}_\times
    \text{, }
    \mathcal{H}^\text{DN}_\text{int} := -\imath
    \begin{pmatrix}
        k_{\mathrm{p}1} & k_{\mathrm{p}2}
    \end{pmatrix}\otimes
    \begin{pmatrix}
        1 \\ 0
    \end{pmatrix}
    \otimes
    \begin{pmatrix}
        \vec{0} & \identity{3}
    \end{pmatrix}
\end{equation*}
and 
\begin{equation*}
    \mathcal{H}^\text{DN}_\text{fluid} := 
    \begin{pmatrix}
        \kappa{-}\delta\kappa & 0 \\ 0 & \kappa{+}\delta\kappa
    \end{pmatrix}\otimes
    \begin{pmatrix}
        0 & \vec{k}^\dagger \\ \vec{k} & \matrix{0}
    \end{pmatrix}\text{ .}
\end{equation*}

With $\vec{k}\eq k\vec{e}_z$ w.l.o.g.\ because of the isotropy of the problem, this double-net eigenproblem yields the longitudinal band
\begin{equation}
    \frac{\omega_l}{c} \approx \sqrt{k_{\mathrm{p}1}^2 + k_{\mathrm{p}2}^2 + \left[1+2\frac{k_{\mathrm{p}2}^2 - k_{\mathrm{p}1}^2}{k_{\mathrm{p}2}^2 + k_{\mathrm{p}1}^2}\frac{\delta\kappa}{\kappa}\right]\kappa^2\,k_l^2}
        \label{eq:long_disp_double}
\end{equation}
with fields $
    \vec{E}\parallel\vec{e}_z\text{ , }
    \vec{H}=0 \text{ , and }
    \vec{j}_i = \parallel \vec{e}_z$
; and the two-fold degenerate transverse light band
\begin{equation}
    \frac{\omega_t}{c} = \sqrt{k_{\mathrm{p}1}^2 + k_{\mathrm{p}2}^2 + k_t^2}
       \label{eq:trans_disp_double}
\end{equation}
with
$    \vec{E}_\pm = \frac{\omega_t}{c}\left(\vec{e}_x{\pm}\imath \vec{e}_y\right) \text{, }
    \vec{H}_\pm = k_t \left(\vec{e}_y{\mp}\imath\vec{e}_x\right)\text{ , }
    V_{i\pm} = 0\text{, and }
    \vec{j}_{i\pm} = \imath k_{\mathrm{p}i}\left(\vec{e}_x{\pm}\imath \vec{e}_y\right) $.
Similar to \secref{sec:single_plasma}, the zero-frequency band is $6$-fold degenerate, with two spuriouos longitudinal solutions and now four transverse current modes with
\begin{align}
\begin{pmatrix}
    H_x \\ H_y
\end{pmatrix} = 
-\frac{k_{\mathrm{p}i}}{k} \sigma_y
\begin{pmatrix}
    j_{ix} \\ j_{iy}
\end{pmatrix}
\end{align}
and all other fields vanishing.

Importantly, a new fundamental band appears below the plasma frequency. It emanates at the $\Gamma$-point at $\omega\eq0$ with constant slope as shown in the main manuscript in Figure 1(b). The associated modes are reminiscent of acoustic waves in a conventional fluid, but with two species of charges fluctuating in opposite directions. The dispersion relation is
\begin{equation}
    \frac{\omega_a}{c} \approx \left( \kappa+\frac{k_{\mathrm{p}1}^2 - k_{\mathrm{p}2}^2}{k_{\mathrm{p}1}^2 + k_{\mathrm{p}2}^2}\delta\kappa\right)\,k_a,
    \label{eq:EAW}
\end{equation}
with fields (in leading order in $\delta\kappa$) $\vec{E}\eq\delta\kappa k\vec{e}_z$, $\vec{H} \eq 0$, $ V_i \eq {-}\imath\kappa({-}1)^i(k_{\mathrm{p}1}^2+k_{\mathrm{p}2}^2)/(4k_{\mathrm{p}i}^2)$ and $\vec{j}_i \eq \frac{k_{\mathrm{p}1}}{\kappa}V_i\vec{e}_z $.
As we can see, the magnetic field of the EAW is strictly zero, while the electric field is longitudinal and linearly depends on both $\delta\kappa$ and $k$.
In the real DNM structures, since the HDP model is only approximately valid, the electric field of the low frequency mode obtains and additional longitudinal term that is zero order in $\delta\kappa$ and quadratic in $k$.
We demonstrate this behavior in \figref{fig:s3} by averaging the fields obtained through full wave simulations for three different DNM geometries.
In \figref{fig:s3} (a) and (b), the $x$-component increases quadratically when moving away from the $\Gamma$-point, while the $y$ and $z$ components vanish for all $\vec{k}\parallel \vec{e}_x$.
In \figref{fig:s3} (c), there is an additional small linear dependence as $\delta\kappa/\kappa\,{\approx}\,0.03$.
In any case, the EAW modes in DNMs are evidently quasi-longitudinal.

\section{Effective Hamiltonian}\label{sec:effective}

The solution of the eigenproblem of the HDP associated with the Hamiltonian in \eqref{eq:EVP2} can be obtained algebraically using any standard computer algebra tool. A manual solution can, however, be more conveniently and elegantly obtained through the application of $\vec{k}\cdot\vec{p}$ perturbation theory at the $\Gamma$-point. Note that this procedure is exact when applied to the EAW band, as we shall see below.

Let us briefly summarize the general approach \cite{dresselhaus2008group}.
We first expand the eigenstates at some $\vec{k}\eq\vec{k}_0{+}\delta\vec{k}$ (at the $\Gamma$-point we have $\vec{k}_0\eq0$) in reciprocal space on the basis of those at $\vec{k}_0$, where $\delta \vec{k}$ has infinitesimal length.
Consider the solution of the Hamiltonian at $\vec{k}_0$ to be known for the eigenspace $\mathcal{U}$ belonging to some eigenvalue $\lambda_0\deq\omega(k_0)/c$.
We thus define the orthonormalized eigenvectors $\vec{u}_\alpha$ that span $\mathcal{U}$, such that
\begin{align*}
    \mathcal{H}(\vec{k}_0)\vec{u}_\alpha=\lambda_0 \vec{u}_\alpha\text{ .}
\end{align*}

If the components of the Hamiltonian are analytical at $\vec{k}_0$, we obtain in first order in $\delta\vec{k}$,
\begin{align*}
    \mathcal{H}(\vec{k}_0{+}\delta\vec{k}) = \mathcal{H}(\vec{k}_0) +
    \underbrace{\delta\vec{k}\cdot\left[\nabla_{\vec{k}}\mathcal{H}\right](\vec{k}_0)}_{\eqd \delta\mathcal{H}}\text{ .}
\end{align*}

At the perturbed point $\vec{k}_0{+}\delta\vec{k}$, the eigenvectors $\vec{v}_n$ are an element of $\mathcal{U}$ in zero order perturbation theory.
This immediately yields the identity (using Einstein notation),
\begin{align*}
    \vec{v}_n =  \left(\vec{u}_\alpha\cdot\vec{v}_n\right) \vec{u}_\alpha
     := c_\alpha^{(n)}\, \vec{u}_\alpha\text{ .}
\end{align*}

As a result, the eigenvalue equation at $\vec{k}_0{+}\delta\vec{k}$ becomes
\begin{align*}
    \left[\lambda_0+\delta\mathcal{H}\right]
    c_\alpha^{(n)}\vec{u}_\alpha =
    \lambda_n c_\alpha^{(n)}\vec{u}_\alpha\text{ .}
\end{align*}

Upon testing with $\vec{u}_\alpha$, we obtain its weak form, which is an algebraic eigenvalue equation that has the dimension of $\mathcal{U}$,
\begin{align}
    \underbrace{\vec{u}_\alpha\cdot\left(\delta\mathcal{H}\vec{u}_\beta\right)}_{\eqd \mathcal{H}_{\alpha\beta}}c_\alpha^{(n)} = 
    \left(\lambda_n-\lambda_0\right) c_\alpha^{(n)}\text{ .}
    \label{eq:Heff}
\end{align}

In other words, the deviation of the $\dim(\mathcal{U})$ eigenvalues $\lambda_n$ from $\lambda_0$ is given by the eigenvalues of the effective Hamiltonian $\mathcal{H}_{\alpha\beta}$, whose entries are the matrix elements of the first order perturbation $\delta\mathcal{H}$ in $\mathcal{U}$.
Since $\mathcal{H}$ is linear in $\vec{k}$ for the HDP Hamiltonian \eqref{eq:Hamiltonian2}, the first order expansion is exact, that is $\mathcal{H}_\text{DN}\eq\mathcal{H}_0{+}\delta\mathcal{H}$ with
\begin{align*}
    \mathcal{H}_0 &= 
    \begin{pmatrix}
        \matrix{0} & \mathcal{H}_{\text{int}} \\
        \mathcal{H}_{\text{int}}^\dagger & \matrix{0}
    \end{pmatrix} \\ \text{and}\quad
    \delta\mathcal{H} &=
    \begin{pmatrix}
        \mathcal{H}_{\text{field}}(\delta\vec{k}) & \matrix{0} \\
        \matrix{0} & \mathcal{H}_{\text{fluid}}(\delta\vec{k})
    \end{pmatrix} \text{ .}
\end{align*}

Since we are interested in the eigenspace of $\mathcal{H}_0$ at vanishing frequency, we are searching for its kernel, which is $8$-dimensional and contains arbitrary $\vec{H}$, arbitrary $V_1$ or $V_2$, and arbitrary $\vec{j}_1\eq{-}\vec{j}_2$, with all other fields vanishing.
Labelling the field components in $\vec{u}_\alpha$ with $i$ and disregarding the trivial electrostatic solution with $\vec{H}\ne0$, we thus obtain the $5$-dimensional eigenspace matrix with orthogonormalized eigenvectors as rows
\begin{align*}
    U_{i\alpha} &= \frac{1}{\sqrt{k_{\mathrm{p}1}^2+k_{\mathrm{p}2}^2}}\\
    &\times\begin{pmatrix}
        0 & 0 & 0 & 0 & 0 \\
        0 & 0 & 0 & 0 & 0 \\
        0 & 0 & 0 & 0 & 0 \\
        0 & 0 & 0 & 0 & 0 \\
        0 & 0 & 0 & 0 & 0 \\
        0 & 0 & 0 & 0 & 0 \\
        \sqrt{k_{\mathrm{p}1}^2+k_{\mathrm{p}2}^2} & 0 & 0 & 0 & 0 \\
        0 & 0 & k_{\mathrm{p}2} & 0 & 0 \\
        0 & 0 & 0 & k_{\mathrm{p}2} & 0 \\
        0 & 0 & 0 & 0 & k_{\mathrm{p}2} \\
        0 & \sqrt{k_{\mathrm{p}1}^2+k_{\mathrm{p}2}^2} & 0 & 0 & 0 \\
        0 & 0 &-k_{\mathrm{p}1} & 0 & 0 \\
        0 & 0 & 0 &-k_{\mathrm{p}1} & 0 \\
        0 & 0 & 0 & 0 &-k_{\mathrm{p}1}
    \end{pmatrix}\text{ .}
\end{align*}

\begin{figure*}[t]
    \includegraphics[width=\textwidth]{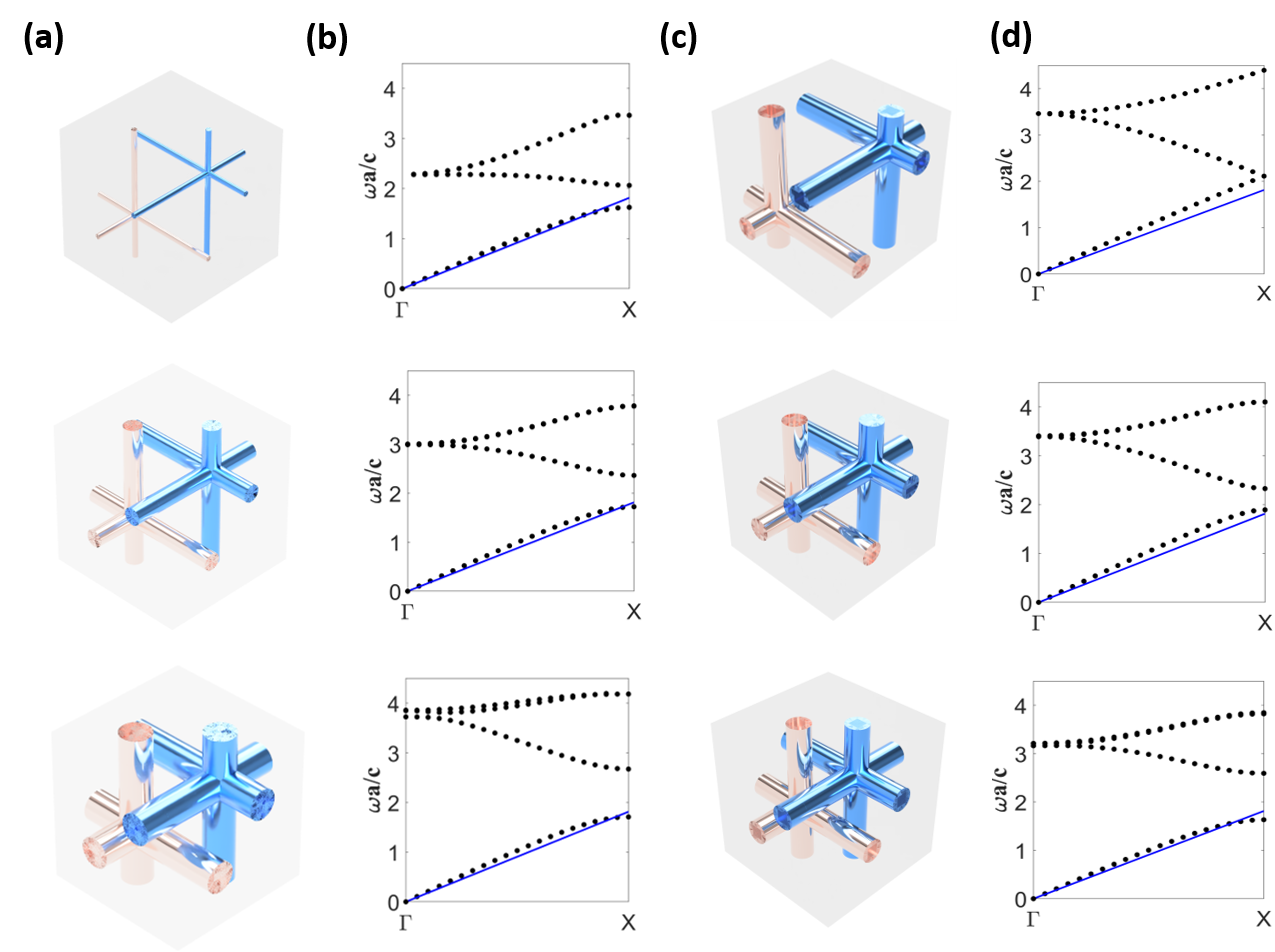}
    \caption{
    (a) Geometries of DNMs with a fixed offset of $a/3(1,1,1)$ of two identical wire meshes.
    The radii are $\SI{0.02}a$ , $\SI{0.06}a$ and $\SI{0.12}a$, respectively; (b) the corresponding bandstructures of (a).
    (c) DNMs with fixed radius of $r\eq\SI{0.08}a$ and varying relative distance $\vec{R}\eq a/2(1,1,1)$, $a/3(1,1,1)$, and  $a/4(1,1,1)$; (d) bandstructures of (c).
    Black dots are full wave simulations, blue lines show the $1/\sqrt{3}$ slope.
    }
    \label{fig:s4}
\end{figure*}

The effective Hamiltonian is thus
\begin{align}
    \mathcal{H}_{\alpha\beta} &= U_{i\alpha}\delta\mathcal{H}_{ij} U_{j\beta} =
    \begin{bmatrix}
    0 &\matrix{h}\\\matrix{h}^\intercal&0
    \end{bmatrix} \text{ ,}\label{eq:Hab}
\end{align}
where
\begin{align*}
    \matrix{h} &= \frac{1}{\sqrt{k_1^2+k_2^2}}\\
    &\times\begin{bmatrix}
    k_{\mathrm{p}1}(\kappa+\delta\kappa) k_x
    &k_{\mathrm{p}1}(\kappa+\delta\kappa) k_y
    &k_{\mathrm{p}1}(\kappa+\delta\kappa) k_z \\
    k_{\mathrm{p}2}(\kappa-\delta\kappa) k_x
    &k_{\mathrm{p}2}(\kappa-\delta\kappa) k_y
    &k_{\mathrm{p}2}(\kappa-\delta\kappa) k_z
    \end{bmatrix} \text{ .}
\end{align*}

Three of the eigenvalues of the perturbation Hamiltonian in \eqref{eq:Hab} are zero. In addition we reproduce the EAW band in \eqref{eq:EAW} and its chirally symmetry counterpart:
\begin{equation*}
    \lambda_\pm \approx \pm \left( \kappa+\frac{k_{\mathrm{p}1}^2-k_{\mathrm{p}2}^2}{k_{\mathrm{p}1}^2+k_{\mathrm{p}2}^2}
    \delta\kappa\right)\,k_a\text{ .}
\end{equation*}

\begin{figure}
    \begin{overpic}[width=\columnwidth]{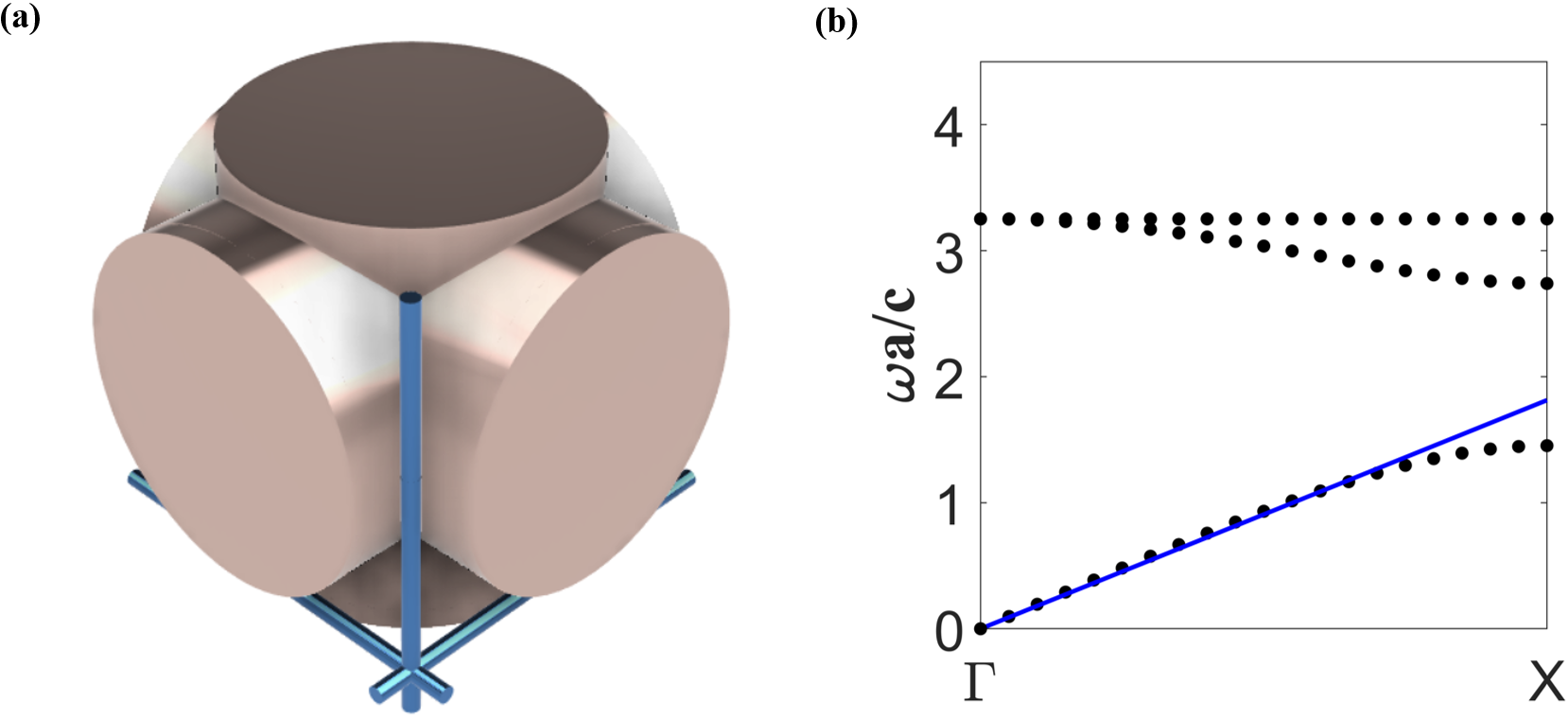}
        \put(0,42){\colorbox{white}{(a)}}
        \put(50,42){\colorbox{white}{(b)}}
    \end{overpic}
    \caption{(a) Extreme DNM with drastically different radii of $\SI{0.36}a$ and $\SI{0.03}a$ for the two nets. The offset is chosen as $0.45a(1,1,1)$; (b) corresponding bandstructure, blue line shows $1/\sqrt{3}$ slope.}
    \label{fig:s5}
\end{figure}
\begin{figure*}
    \includegraphics[width=\textwidth]{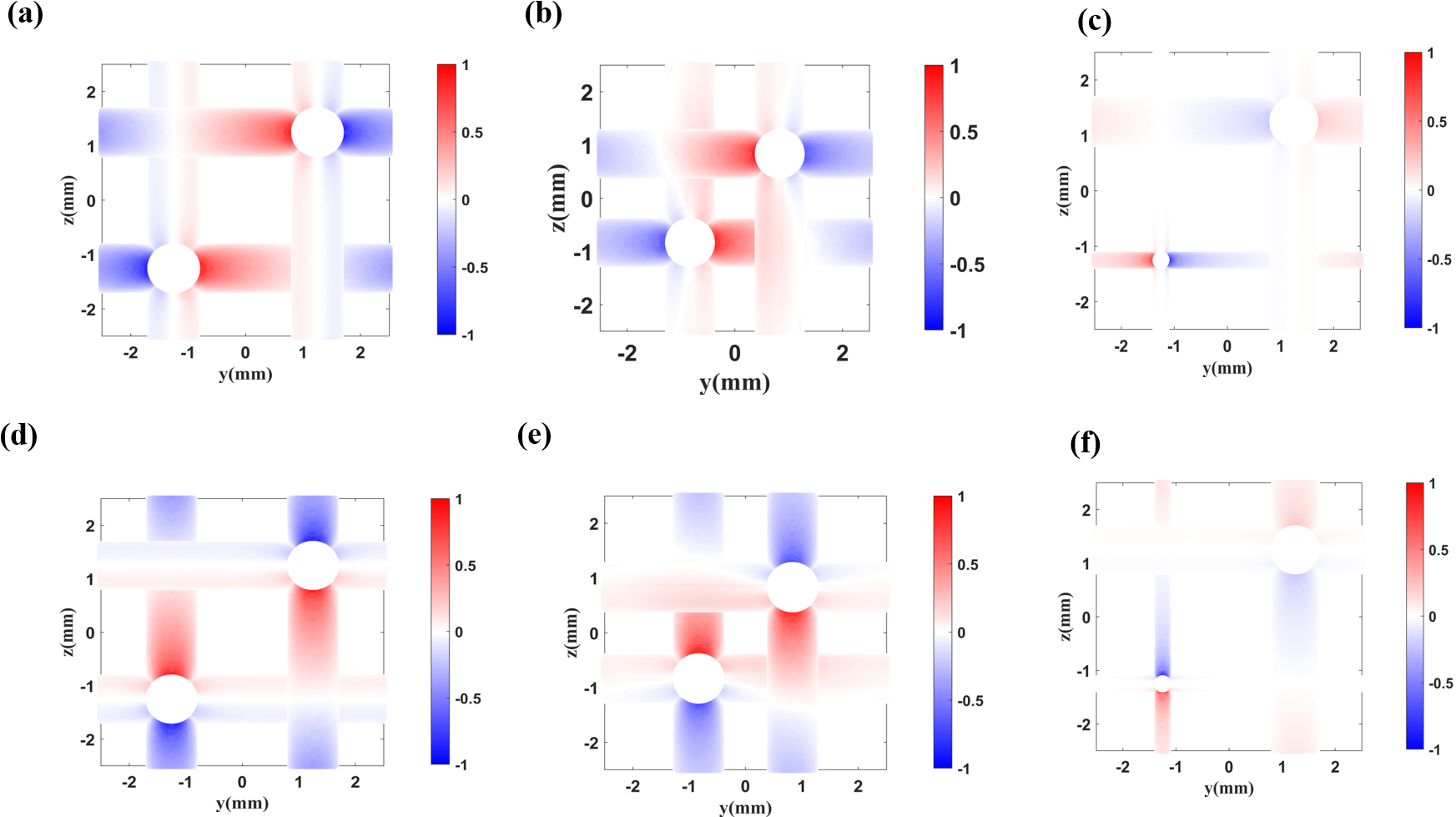}
    \caption{$y$ and $z$ components of surface current densities $J_s$ of DMNs with different parameters near the static limit at $k_{x}\eq 0.05 \frac{\pi}{a}$ and $k_y\eq k_z\eq0$. (a), (b)and (c) show the $J_{sy}$  distribution, while (b), (d) and (f) show $J_{sz}$. DNM parameters  for (a) and (d) are $(0.08~,0.08~,0.5)$; for (b) and (e) $(0.08~, 0.08~,0.33)$; for (c) and (f) $(0.08~, 0.02~,0.5)$.}
    \label{fig:s6}
\end{figure*}

Evidently, the EAW slope only weakly depends on the difference in plasma frequencies of the two networks in first order $\delta\kappa$. Particularly, if the two networks have equal radius, the slope of the EAW band is fully determined by the pressure parameter $\kappa$ only. Generally, we note that the dispersion of EAW band in the low frequency limit is robust against changes in the DNM parameters such as the radius of the metal wires and the relative offset between the two nets. This assumption is verified by simulating DNMs with various parameters, shown in \figref{fig:s4} and discussed in \secref{sec:effective}.  

\section{Bandstructure of DNMs with different parameters}\label{sec:params}

\begin{figure*}
    \begin{flushleft}
    \begin{overpic}[width=.47\textwidth]{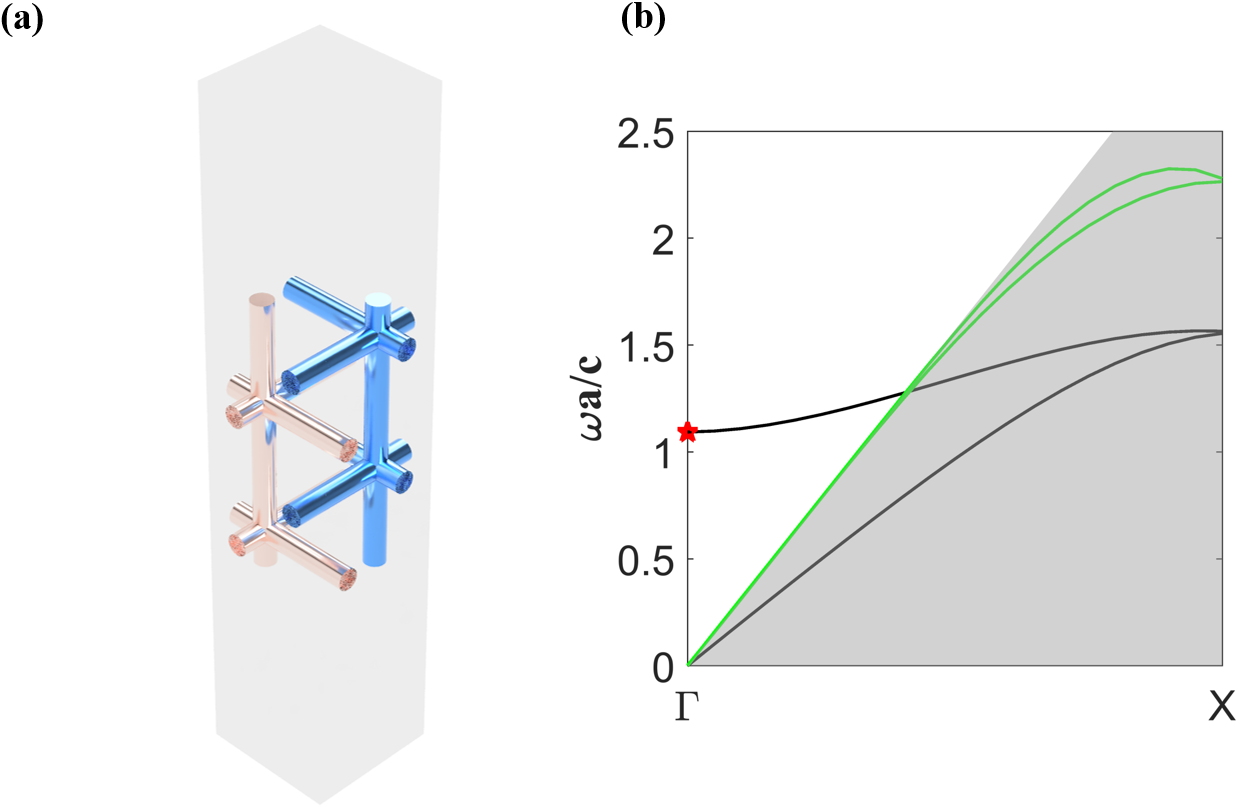}
    \put(110,0){\includegraphics[width=.47\textwidth]{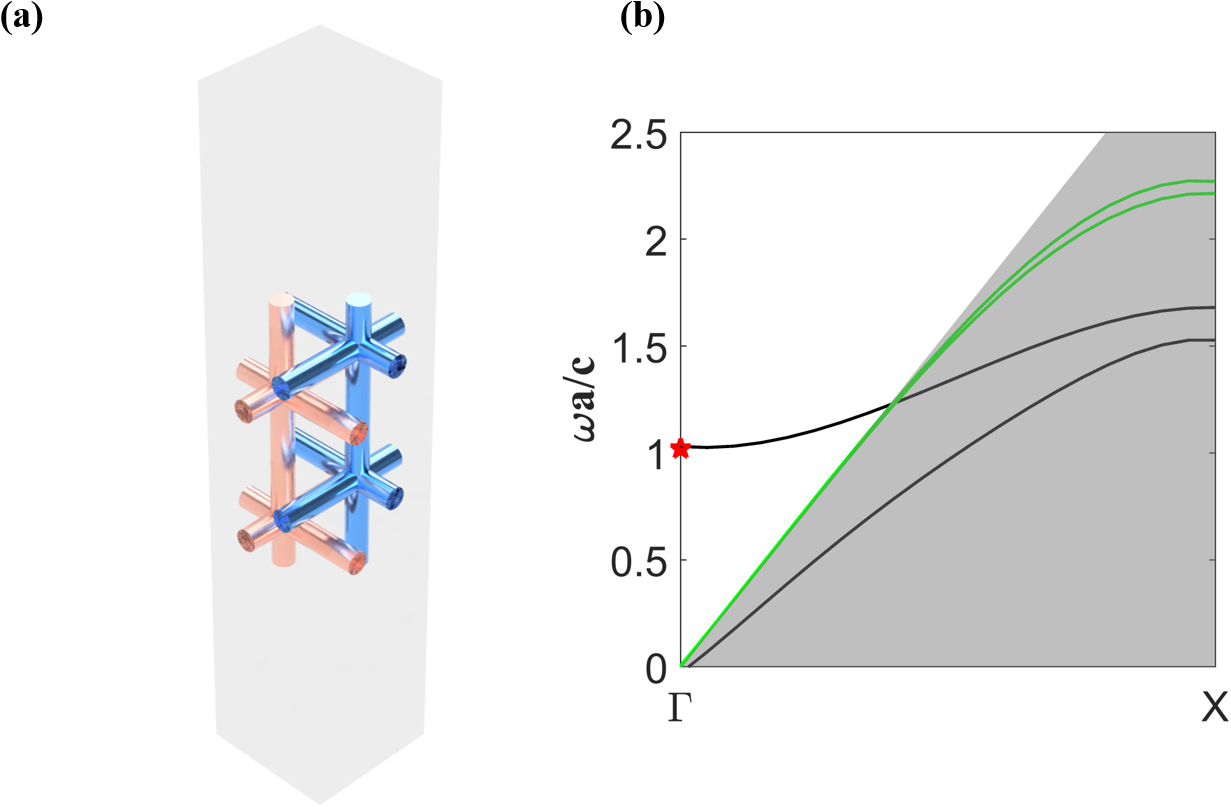}}
    \put(0,-90){\includegraphics[width=.47\textwidth]{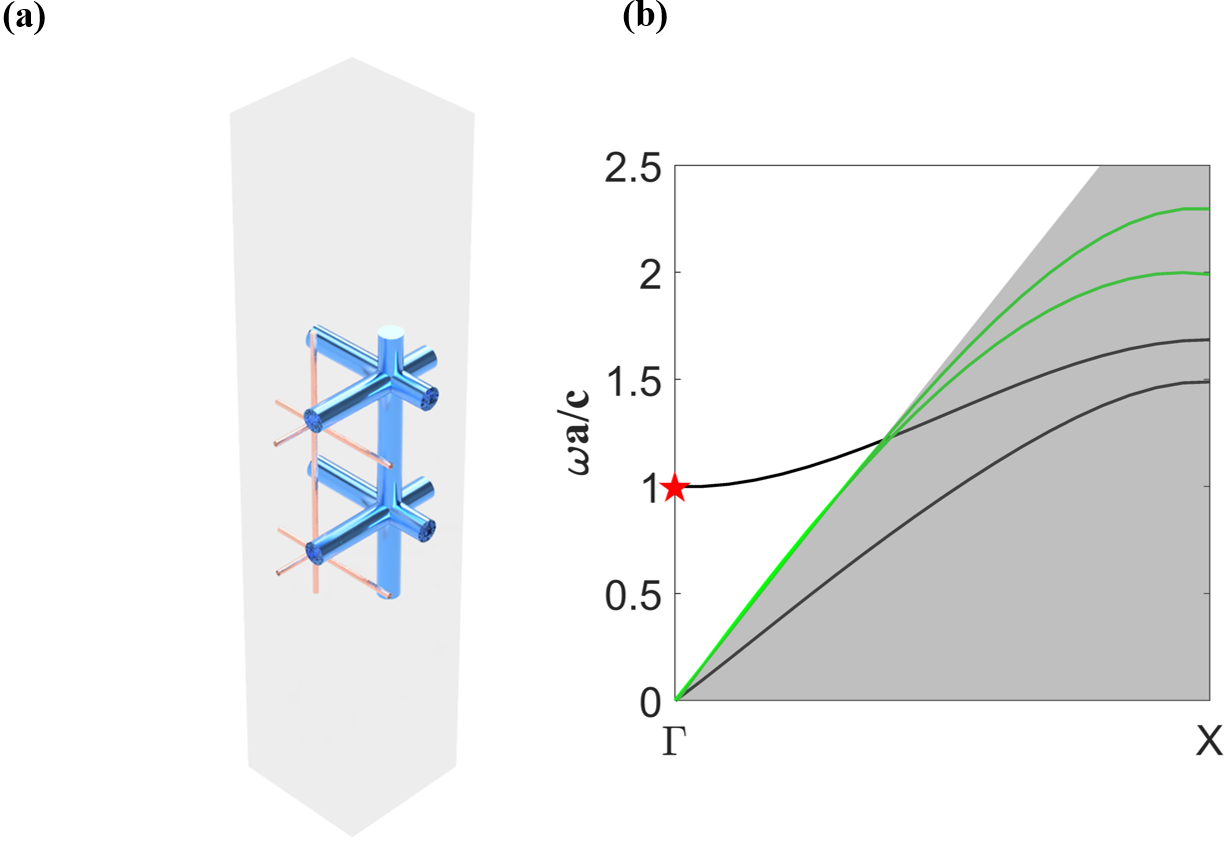}}
    \put(110,-100){\includegraphics[width=.47\textwidth]{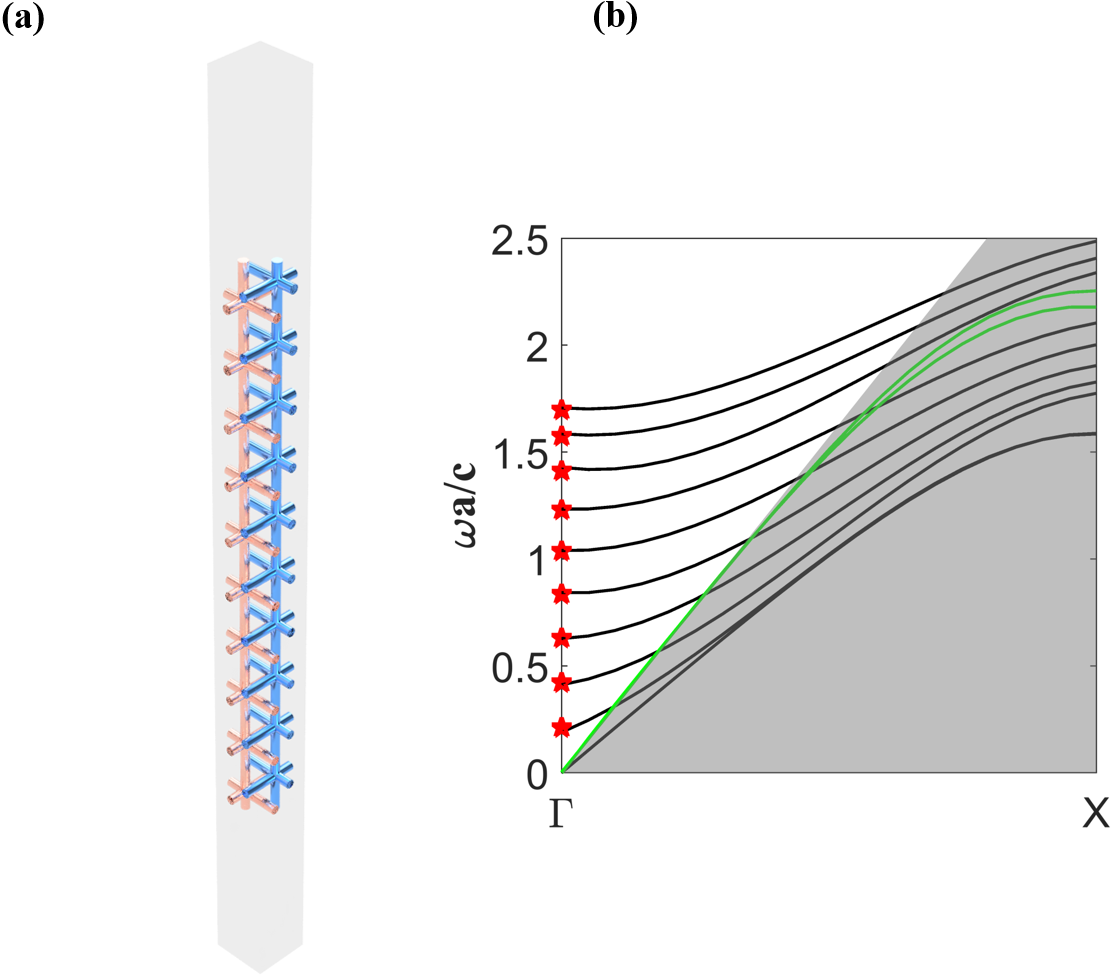}}
    \put(0,62){\highlight{(a)}}
    \put(50,62){\highlight{(b)}}
    \put(106,62){\highlight{(c)}}
    \put(156,62){\highlight{(d)}}
    \put(0,-15){\highlight{(e)}}
    \put(50,-15){\highlight{(f)}}
    \put(0,-23){\highlight{\color{white}{(e)}}}
    \put(50,-23){\highlight{\color{white}{(f)}}}
    \put(106,-15){\highlight{(g)}}
    \put(160,-15){\highlight{\color{white}{(h)}}}
    \put(155,-15){\highlight{(h)}}
    \end{overpic}
    \end{flushleft}\vspace{25em}
    \caption{(a) Geometry of a DNM slab consisting of two identical nets with $\vec{P}\eq(2/25,2/25,1/2)$.
    The slab thickness is $2a$.
    (b) Bandstructure of the slab shown in (a).
    The black band emanating at zero frequency is made by an in-plane EAW mode, while one quasi-BIC band emanates from the BIC mode indicated by a red star.
    The green bands are spoof plasmon surface modes confined to the slab surface.
    (c) Slab with tickness $2a$ and $\vec{P}\eq(2/25,2/25,1/3)$.
    (d) Bandstructure for the slab shown in (c).
    The BIC mode and the associated quasi-BIC band are conserved under the symmetry break.
    (e) Slab with thickness $2a$ and $\vec{P}\eq(2/25,1/50,1/3)$.
    (f) Bandstructure for the slab shown in (e).
    (g) Slab with thickness $10a$ and $\vec{P}\eq(2/25,2/25,1/3)$.
    (h) Bandstructure for the slab shown in (g).
    As theoretically predicted, $9$ BIC modes are found 
    }
    \label{fig:s7}
\end{figure*}

In \secref{sec:HDP}, we homogenized the DNM with an HDP model. We established that the EAW mode in the DNM is equivalent to two species of charged particles moving in opposite directions. From this model, we further obtained an exact dispersion relation \eqref{eq:EAW} in the long wavelength limit, which mainly depends on the average pressure parameter of the two networks, which is $\kappa\gtrsim 1/\sqrt{3}$. To verify this statement,  we numerically simulate DNMs with different radii (\figref{fig:s4}(a)-(b)) and relative positions (in \figref{fig:s4}(c)-(d)).

In \figref{fig:s4}(a) we begin with two wire meshes with a fixed relative space offset $a/3(1,1,1)$ and different plasma frequency tuned by the (equal) radius of the two wire meshes from  $\SI{0.02}a$, over $\SI{0.06}a$, to $\SI{0.12}a$. The corresponding bandstructures are shown in \figref{fig:s4}(b). The black dots are from full wave simulations, while the blue solid lines show the limiting $\omega/c\eq\frac{1}{\sqrt{3}} k$ dispersion. As expected, the DNM slope in the vicinity of the $\Gamma$-point is larger in all three cases, and approaches the limit for small radii.
The model is of course invalid when approaching the $X$-point, due to the intrinsic  non-homogeneous and periodic nature of the DNM structures.
In \figref{fig:s4}(c), we demonstrate DNMs with various offset while keeping the plasma frequency of both nets equal and constant (the radius is $\SI{0.08}a$). The bandstructures are shown in \figref{fig:s4}(d). The results are again consistent with the theoretical expectation from the HDP model. Upon close inspection, the low frequency slope approaches the blue limit, when the two networks come closer together.

To further verify the validity of the HDP model applied to DNMs, we demonstrate an extreme case In \figref{fig:s5}  with a huge difference in radius of the two nets.

\section{Quadrupole-like currents in DNMs with different parameters}\label{sec:SI4} 

The surface current densities of the two nets in DNMs parallel to the propagation direction, with opposite signs, resulting in a vanishing homogenized electromagnetic field when approaching the $\Gamma$-point, as shown in Fig.~2 in the main text.
Meanwhile, small currents flow in the wires oriented perpendicularly to the propagation direction, which we refer to as $x$ w.l.o.g., at finite frequencies.
The pcu-c DNM with body centered cubic symmetry has a global $C_{4v}$ symmetry through the center of each of the wires.
Since the currents in the $x$ wires are constant along the wire surface, they transform trivially under the corresponding $C_{4v}$ symmetry.
As a consequence, the weak EAW currents in the wires in the $y$-$z$ plane must transform trivially as well, and therefore either all flow towards the nodes or away from them in a $4$-fold quadrupole-like fashion.
This expected behavior is demonstrated through the surface current densities obtained from full wave simulations, whose $y$ component is shown in \figref{fig:s6} (a) and the $z$ component in \figref{fig:s6} (d).
\begin{figure}
    \includegraphics[width=0.5\textwidth]{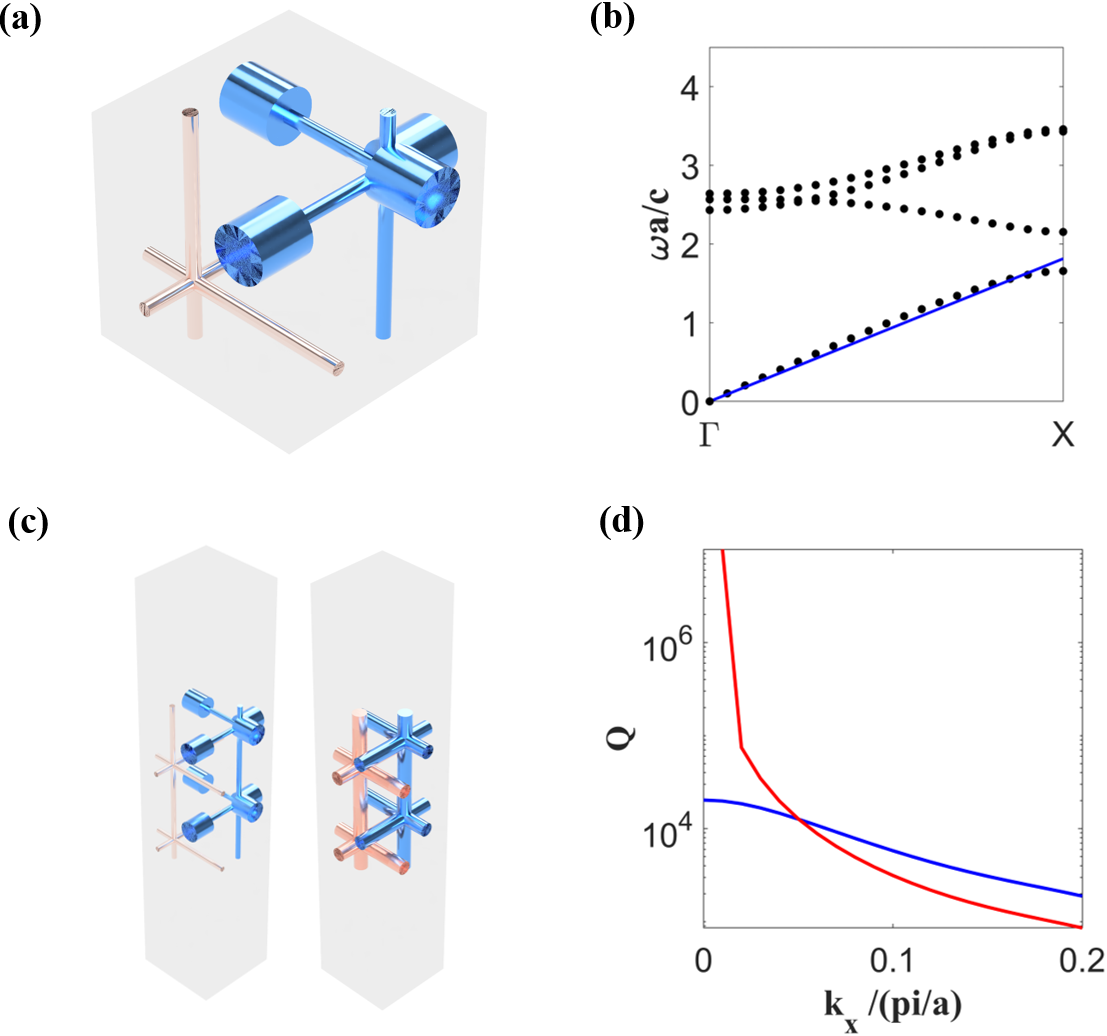}
    \caption{(a) Schematic diagram of a DMN with broken local $C_{4v}$ symmetry; the radius of the added thick metal cylinder is $0.12a$, while the radius of the unperturbed net is $0.02a$.
    (b) Corresponding dispersion relation of (a); the black dots are from a full wave simulation, the blue line is the $1/\sqrt{3}$ slope.
    (c) DMN slab with (left) and without (right) breaking the local symmetry.
    (d) $Q$-factor of the nets shown in (c). The $Q$-factor diverges at the $\Gamma$-point for the unperturbed network (red line), which is indicative of the existence of BICs.
    In contrast, the $Q$-factor  stays finite for the network with strong local symmetry breaking (blue), showing quasi-BIC behavior only.}
    \label{fig:s8}
\end{figure}
\begin{figure}
    \includegraphics[width=\columnwidth]{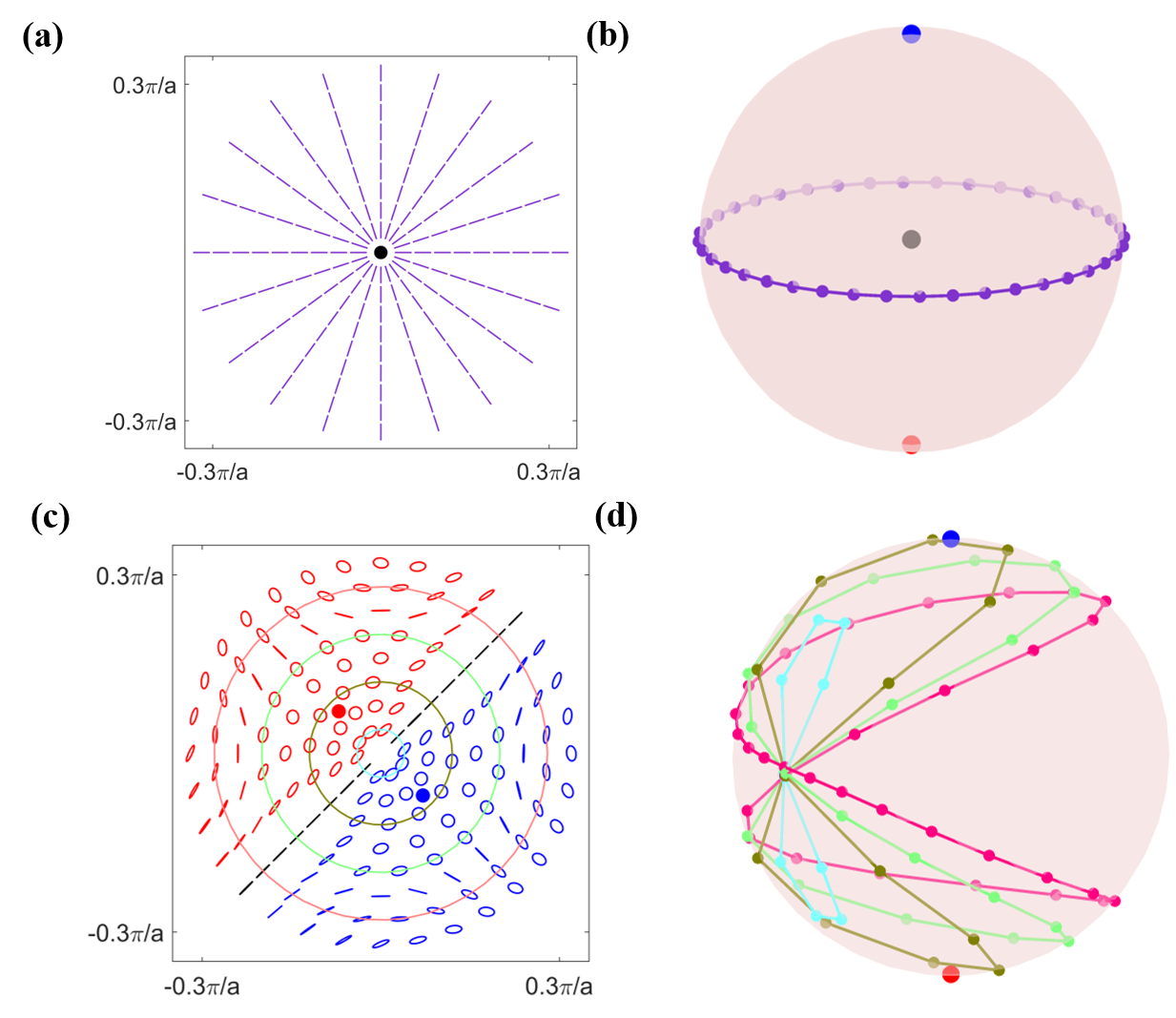}
    \caption{(a) 2D far field polarization pattern of BICs in DNMs.
    (b) Polarization along a circle in (a), projected onto the Poincaré sphere.
    (c) Far field polarization pattern of the quasi-BIC for the perturbed DNM.
    (d) Polarization projected onto the Poincaré sphere along the circles shown in (c).}
    \label{fig:s9}
\end{figure}

Notably, the in-plane geometry of the individual nets dominates the electromagnetic distribution, since the currents on the two nets experience very weak inter-coupling at low frequencies.
The local $C_{4v}$ symmetry of each single net thus guarantees the formation of the aforementioned quadrupole-like current distributions, even in the absence of a global $C_{4v}$ symmetry.
We demonstrate this behavior through the current distributions shown in \figref[(b)]{fig:s6} and (e) for DNMs with a non-symmetric shift, and in \figref[(c)]{fig:s6} and (f) for distinct wire radii.
To a very good approximation, these fields retain their trivial character with respect to the individual net's $C_{4v}$ symmetry.
Importantly, this trivial character is evidently incompatible with the free space photon modes, which carry the non-trivial character of the $E_\pm$ (right and left circular polarization) irreducible representation \cite{PhysRevB.88.245116}.

\section{BICs in the DNM immune to global symmetry breaking }\label{sec:SI5}

The in-plane cross-shaped geometry which intersects perpendicularly by identical metal pillar of each individual wire mesh keeps the $C_{4v}$ symmetry. The cross-shaped metallic pillar guide the surface density current, forming two quadruples in the space. The coupling between the two electrical quadruples is weak which is also the mechanism causing the BICs modes in DNM slab to be immune to the global symmetry. In the main text, we show the DNM slab with the parameters R(0.4,0.4) and offset (a/3, a/3,a/3). To further illustrate robustness of BICs in  DNM, we show numerical results of DNM slabs with various parameters in \figref{fig:s7}, all of which host BIC states.

\section{Lifting BICs by breaking local symmetry}\label{sec:SI6}
 
Throughout this work, we have considered the metals constituting the DNMs as perfect electric conductors, so that surface currents flow on the metallic wires and determine the microscopic electromagnetic modes.
We have shown in \secref{sec:SI5} that the surface current density  forms two quadrupole-like distributions, which are robust against global symmetry breaking.
These quadrupole currents are in the 1D representation of the $C_{4v}$ point group labelled $B_1$ in \cite{dresselhaus2008group}, and thus cannot couple to vacuum radiation, which is in the 2D $E$ representation.
However, the BICs can be destroyed by breaking the local $C_{4v}$ symmetry of the individual nets in the DNM.
We demonstrate this by introducing an thick metal cylinder segment  on the lateral wires along the $x$ and $y$ directions normal to the slab direction $z$ (\figref{fig:s8}(a)).
These additional cylinders are arranged in such a way that the $C_{4v}$ symmetry is reduced to a single mirror plane on the $x{-}y$ diagonal.
For simplicity, we leave the other net unperturbed.
The broken local symmetry, reduced to merely a mirror group, enables couplings of the EAW mode to vacuum modes.
When compared to  the DNM with two identical unperturbed wire meshes, the slab formed by the perturbed DNM (\figref{fig:s8}(c)) degrades the BIC modes at the $\Gamma$-point to quasi-BICs with finite $Q$-factors (\figref{fig:s8}(d)).
It is, however, worth noting that the perturbed DNM slab can still be well described by the HDP theory, as illustrated by the
bulk bandstructure along the out-of-plane $z$ direction in \figref{fig:s8}(b), which retains the typical EAW band.

The nature of the quasi-BICs, on the other hand, necessarily imposes changes in polarization structure in the far field.
Compared with the `star' polarization pattern of the BICs in the unperturbed structure (\figref{fig:s9}(a)), the perturbed DNM slab exhibits a richer polarization distribution, forming two patches of elliptically polarized states with opposite handedness, separated by a diagonal linear polarization line (\figref{fig:s9}(c)).
The position of this line on the diagonal is evident from the fact that the in-plane components of both the dominating longitudinal currents and the quadrupole currents are even with respect to the remaining mirror if the mirror coincides with the plane of incidence, that is, if the wave vector is invariant.
Thus, the far-field must obey the same symmetry classification, that is, it must be $p$ polarised, with the electric field confined to the plane of incidence.
In all other directions, the EAW field is a combination of even (quadrupole currents) and odd (in-plane longitudinal currents) contributions, and thus the far-field is generally elliptically polarized.

The mirror plane divides the $x{-}y$ plane into two half-planes of opposite elliptical polarizations, which is immediately evident from the action of mirror symmetry $\sigma_d$ on the modes on one half-plane.
The vortex at the origin splits into two circular polarization points on either side.
This mechanism is similar to the one reported in \cite{Liu2019}; it is of topological origin.
The appearance of the circular polarization points becomes evident from tracing the polarization on various momentum space circles of different raddi centered around the $\Gamma$-point (\figref{fig:s9}(c)).
The polarization path projected on the Poincaré sphere goes twice through the linear polarization state on the mirror plane and unfolds from a small figure-of-eight (turquoise, small radius) towards a double-loop around the equator (grey, large radius), and thus necessarily passes through the north/south poles (\figref{fig:s9}(d)).

\figref{fig:s9} further indicates that the circular polarized points lie on the axis perpendicular to the remaining mirror plane.
We can understand this behavior from a symmetry perspective. Consider the action of the time-reversal operator $\tau$, which transforms the slab mode with wave vector $\vec{k}_\parallel$ into one with ${-}\vec{k}_\parallel$ and complex conjugated field, that is $\tau \vec{E}_{\vec{k}_\parallel}\eq\vec{E}_{{-}\vec{k}_\parallel}^*$.
The mirror symmetry on the other hand interchanges the $x$ and $y$ components of the field and also maps $\vec{k}_\parallel$ to its negative.
The combined operation $\tau\sigma_d$ thus leaves $\vec{k}_\parallel$ invariant on the line perpendicular to the mirror plane.
We thus obtain $E_x\overset{!}{\eq}E_y^*$, so that the normalized in-plane electric far-field depends on one free parameter $\varphi$ only,
\begin{equation*}
    \vec{E}_\parallel = \frac{1}{\sqrt{2}}
    \begin{pmatrix}
        e^{\imath\varphi} \\ 
        e^{-\imath\varphi}
    \end{pmatrix}\text{ .}
\end{equation*}

In other words, the field stays on a line of constant longitude on the Poincar\'e sphere, connecting zero linear polarization along the mirror plane (at $\Gamma$) and perpendicular to it (away from $\Gamma$ for small perturbations).
It must therefore pass the circular polarized poles ($\varphi\eq\pi/4$) on either side of the mirror plane.

\section{\label{sec:analytical} Analytical quasi-BICs and transmission spectra}

We here devise an analytical model that reveals the connection between the EAW bulk picture and the slab modes. It generates a good prediction for the quasi-BIC modes and the transmission spectra.

Let us first solve the scattering problem using two counter-propagating EAW slab modes only.
We first define the two vacuum domains \Romannum{1} and \Romannum{3}, semi-infinite in $z$-direction and infinite in $x$ and $y$, and the DNM slab domain \Romannum{2}.
As the electromagnetic fields vanish in the HDP model, we need to extract fields at the domain interfaces from the simulations.
The EAW fields in the DNM exhibit longitudinal electric fields and vanishing magnetic fields, as we have already seen in \figref{fig:s3}.
To solve the homogenized Maxwell equations (which is equivalent to cutting the Rayleigh series at the fundamental Bragg order) at the slab surface, however, we need to average the fields over the this surface rather than the whole unit cell \cite{cryst5010014}. Considering an in-plane wave vector $q$ along the $x$ direction, we obtain a homogenized electric field pointing along $x$, and a magnetic field pointing along $y$. This behaviour is consistent with the symmetry classification for the mirror symmetric DNM, and means that the EAWs only couple to $p$ polarised waves in the vacuum domains.
One further obtains that the surface impedance $Z\deq E_x/H_y$ is largely independent of the frequency, and, consistent with optical reciprocity, quadratically depending on $q$ in good approximation.
For an EAW wave propagating in positive $z$ direction, we obtain
\begin{align*}
    Z_2 \approx 2.5 \left(\frac{qa}{2\pi}\right)^2 \text{ .}
\end{align*}
The vacuum impedance in contrast is of course frequency dependent and finite at vanishing $q$, with
\begin{align*}
    Z_1 = \sqrt{1-\left(\frac{cq}{\omega}\right)^2}
\end{align*}
for a wave propagating in positive $z$ direction.

We assume a slab of thickness $d\eq Na$ and the wave numbers in propagation direction are in vacuum $k_1\deq\sqrt{(\omega/c)^2-q^2}$, and in the slab $k_2\deq\sqrt{[\omega/(\kappa c)]^2-q^2}$, using \eqref{eq:EAW} for equal wire thickness of $0.08 a$ ($\kappa\,{\approx}\,0.72$). The non-vanishing field components for the scattering problem can thus be expressed as
\begin{subequations}
\begin{align}
    H_\text{\Romannum{1}} &=\left(e^{\imath k_1(z+d/2)} + r e^{-\imath k_1(z+d/2)}\right) e^{\imath q x} \text{ ,}\\
    E_\text{\Romannum{1}} &= Z_1\left(e^{\imath k_1(z+d/2)} - r e^{-\imath k_1(z+d/2)}\right) e^{\imath q x}\text{ ,}\\
    H_\text{\Romannum{2}} &=\left(A e^{\imath k_2 z} + B e^{-\imath k_2 z}\right) e^{\imath q x} \text{ ,}\\
    E_\text{\Romannum{2}} &= Z_2\left(A e^{\imath k_2 z} - B e^{-\imath k_2 z}\right) e^{\imath q x}\text{ ,}\\
    H_\text{\Romannum{3}} &= t e^{\imath k_1(z-d/2)} e^{\imath q x} \text{ ,}\\
    E_\text{\Romannum{3}} &= Z_1 t e^{\imath k_1 z} e^{\imath q x}\text{ .}\\
\end{align}
\end{subequations}

\begin{figure}
    \includegraphics[width=\columnwidth]{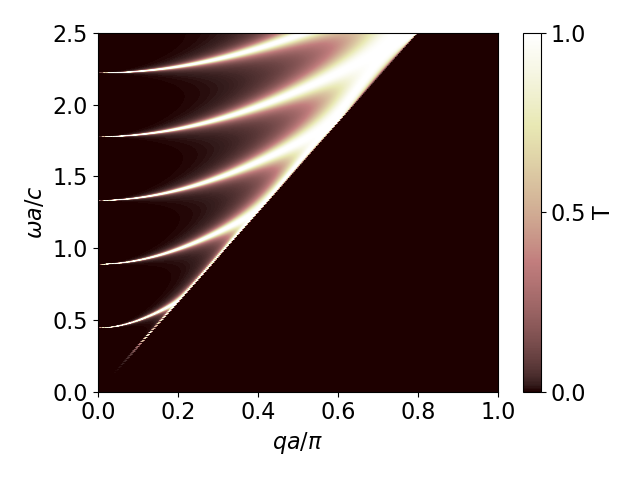}
    \caption{Analytical transmissivity through an $N\eq5$ DNM slab using the HDP dispersion \eqref{eq:EAW} and a homogenized DNM impedance.}
    \label{fig:analytical_T}
\end{figure}

The homogenized Maxwell equations are satisfied if the fields match at both interfaces at $z\eq{\pm}d/2$, yielding a standard scattering problem
\begin{align}
    \begin{pmatrix}
        Z_1 & 0 & Z_2p_2 & -Z_2/p_2 \\
        0 & -Z_1 & Z_2/p_2 & -Z_2p_2 \\
        -1 & 0 & p_2 & 1/p_2 \\
        0 & -1 & 1/p_2 & p_2
    \end{pmatrix}
    \,
    \begin{pmatrix}
        r \\ t \\ A \\ B
    \end{pmatrix} = 
    \begin{pmatrix}
        Z_1 \\ 0 \\ 1 \\ 0
    \end{pmatrix},
\end{align}
 with $p_2\deq \exp\{{-}\imath k_2 d/2\}$. The resulting transmissivity $T\eq|t|^2$ is shown in \figref{fig:analytical_T}. It agrees well with the numerical full wave results in Fig.~3 (d) in the main manuscript, except for a small region close to the light line. Here, the transmission is overestimated significantly, as the surface field homogenization breaks down, particularly at large frequencies.

The same fundamental idea can be employed to calculate the quasi-BIC bands.
For this, we consider only outgoing waves in the vacuum domains, which transport energy away from the slab, that is, with $\Re\{k_1\}{<}0$ [${>}0$] in region \Romannum{1} [\Romannum{3}].
We can characterize the quasi-normal modes as even or odd regarding the $z\mapsto{-}z$ mirror symmetry of the homogenized slab, and therefore substantially reduce the complexity of the problem.
In particular, one only needs to consider the fields in one vacuum domain and the even or odd superposition of EAW modes in the slab.
The fields are thus expressed by
\begin{subequations}
\begin{align}
    H_\text{\Romannum{1}} &= A e^{-\imath k_1(z+d/2)} e^{\imath q x} \text{ ,}\\
    E_\text{\Romannum{1}} &= -Z_1 e^{-\imath k_1(z+d/2)} e^{\imath q x}\text{ ,}\\
    H_\text{\Romannum{2}} &=\left( e^{\imath k_2 z} \pm e^{-\imath k_2 z}\right) e^{\imath q x} \text{ ,}\\
    E_\text{\Romannum{2}} &= Z_2\left( e^{\imath k_2 z} \mp e^{-\imath k_2 z}\right) e^{\imath q x}\text{ .}
\end{align}
\end{subequations}

For convenience, we have here normalized the eigensolutions such that the positive EAW wave in the slab carries a coefficient of $1$.
The matching conditions at the slab surface thus yield
\begin{subequations}
\begin{align}
    A &= p_2 \pm \frac{1}{p_2}\label{eq:Hcont} \\
    -Z_1 A &= Z_2 \left( p_2 \mp \frac{1}{p_2}\right)\text{ .}\label{eq:Econt}
\end{align}
\end{subequations}

At the $\Gamma$-point ($q\eq0$), we immediately obtain $A\eq0$ from \eqref{eq:Econt}. This implies a modal solution with vanishing homogenized field outside of the slab and thus infinite quality factor, in other words BICs.
\eqref{eq:Hcont} further yields $p_2^2\eq\mp1$, which reproduces the BIC frequencies obtained from the HDP model under hard wall boundary conditions, that is,
\begin{equation*}
    \frac{\omega a}{c} = \frac{\kappa \pi n}{N}\quad (n\in\mathbb{N})\text{ .}
\end{equation*}

For arbitrary $q$, \eqref{eq:Hcont} and \eqref{eq:Econt} yield a finite $A$ and are transcendental in the complex frequency.
They therefore need to be solved numerically.
We divide \eqref{eq:Econt} by $Z_1{\ne}0$ and add it to \eqref{eq:Hcont} to obtain
\begin{equation*}
    f(\omega) = \left(Z_1+Z_2\right)p_2
    \pm \left(Z_1-Z_2\right)/p_2 = 0\text{ .}
\end{equation*}

\begin{figure}
    \includegraphics[width=\columnwidth]{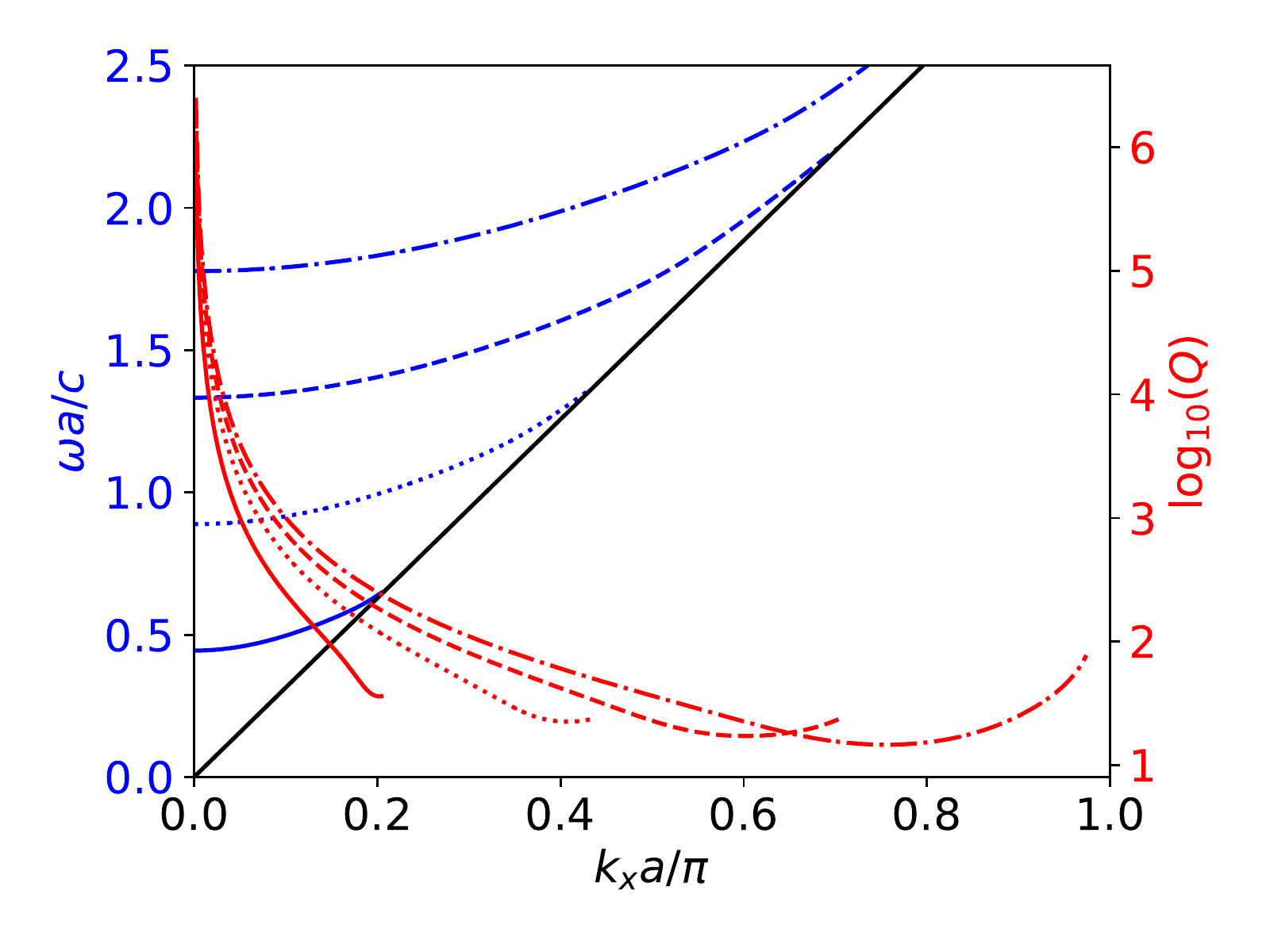}
    \caption{Bandstructure and quality factor of the quasi-BIC bands for a $N\eq5$ DNM slab using the HDP dispersion \eqref{eq:EAW} and a homogenized DNM impedance.}
    \label{fig:analytical_QNM}
\end{figure}

Since $f(\omega)$ is holomorphic, except when crossing the light line, we employ a standard Newton method using $\omega_n$ as starting guess for $q_{n+1}\eq q_n{+}\delta q$ ($n\,{\in}\,\mathbb{N}$) to obtain the  quasi-BIC bands and corresponding quality factors $Q\deq \Re\{\omega\}/(2\Im\{\omega\})$ in \figref{fig:analytical_QNM}.
As expected, the bandstructure and quality factors predict the position and width of the transmission bands in \figref{fig:analytical_T}, respectively.

\begin{figure*}
    \includegraphics[width=\textwidth]{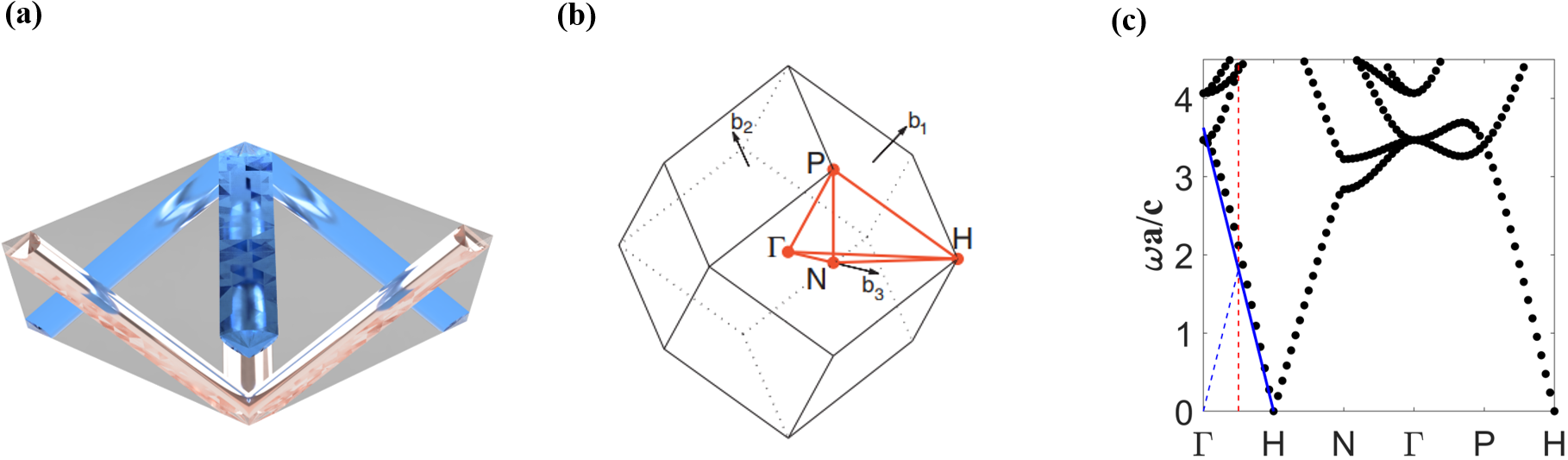}
    \caption{(a) A primitive unit cell of the pcu-c DNM with BCC symmetry. The radius of the two nets is $ r_1\eq r_2\eq 0.08 a$. (b) BCC Brillouin zone. (c) Band diagram of the DNM in (a). The black dots are calculated by full wave simulations, the blue solid line is the $1/\sqrt{3}$ HDP predication. The dashed line illustrates the band folding into the simple cubic Brillouin zone.  }
    \label{fig:s10}
\end{figure*}

\begin{figure*}
    \begin{minipage}[t]{0.48\textwidth}
        \begin{overpic}[width=\textwidth]{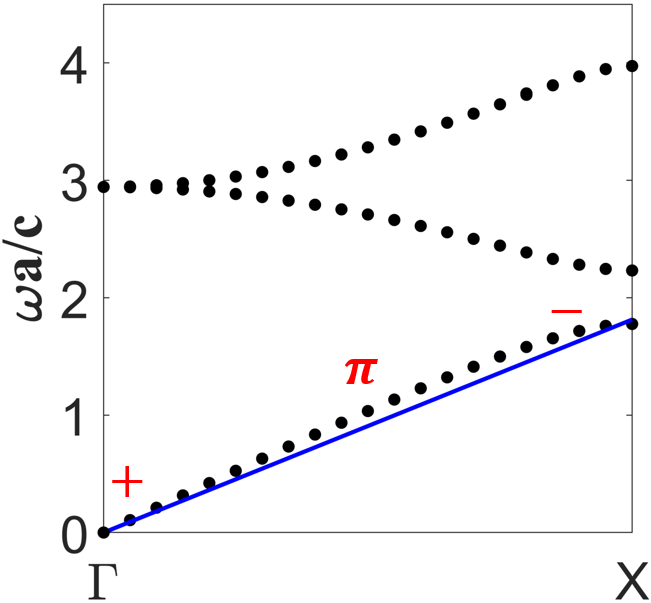}
        \put(0,90){(a)}
        \end{overpic}
    \end{minipage}
    \hfill
    \begin{minipage}[t]{0.48\textwidth}
        \begin{overpic}[width=\textwidth]{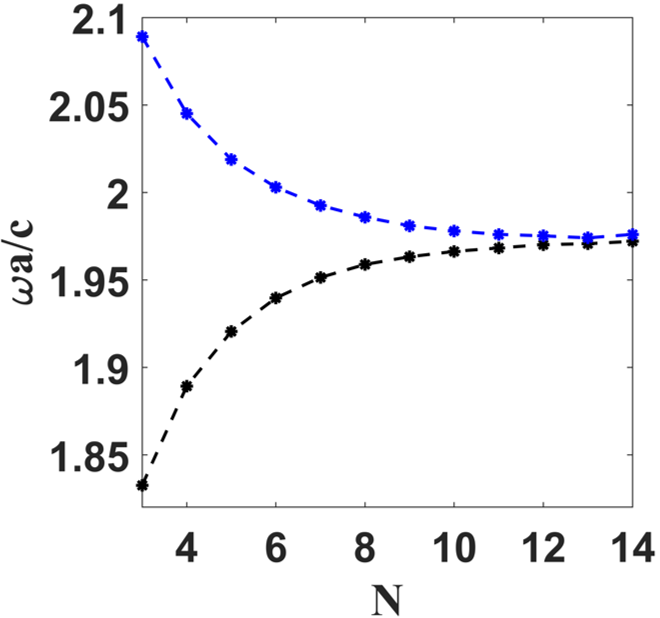}
        \put(0,90){(b)}
        \end{overpic}
    \end{minipage}
    \caption{(a) Bulk bandstructure of the $\vec{P}\eq(2/25,1/50,1/2)$ DNM along $\Gamma{-}X$. Black dots are from full wave simulations, the blue line is the $1/\sqrt{3}$ HDP slope.
    The parity is even [odd] at the high symmetry points $\Gamma$ [$X$], indicated by '$+$' and '$-$', yielding a non-trivial Zak phase of $\pi$ for the unit cell with the thick wire in the center.
    (b) Frequency splitting of the TSBICs at the $\Gamma$-point due to hybridization of the two surface states as a function of slab thickness given by the number of unit cells $N$.
    }
    \label{fig:s11}
\end{figure*}

\section{\label{sec:SI7}Negative refraction in DNMs with BCC symmetry}
The DNM modes are naturally Bloch waves due to the discrete periodicity of the geometry.
Crystal symmetries in the DNM  play an important role and generate additional physics that cannot be predicted by a homogenization theory such as the HDP model.
The crystal symmetry can for example lead to band degeneracies at high symmetry points in the Brillouin zone, where the wave vector is mapped onto itself or a reciprocal lattice equivalent under point group operations.
If a non-primitive unit cell is chosen, artificial band folding, or better a band translation into the artificially decreased Brillouin zone, occurs.
This can have a significant effect on the physical interpretation of the modes as this back-translation can ostensibly change the sign and magnitude of the phase velocity, while keeping the slope of the band and thus the group velocity of the mode unchanged.
One might therefore expect left-handed propagation where the mode is right-handed and vice versa.

\begin{figure*}
    \includegraphics[width=\textwidth]{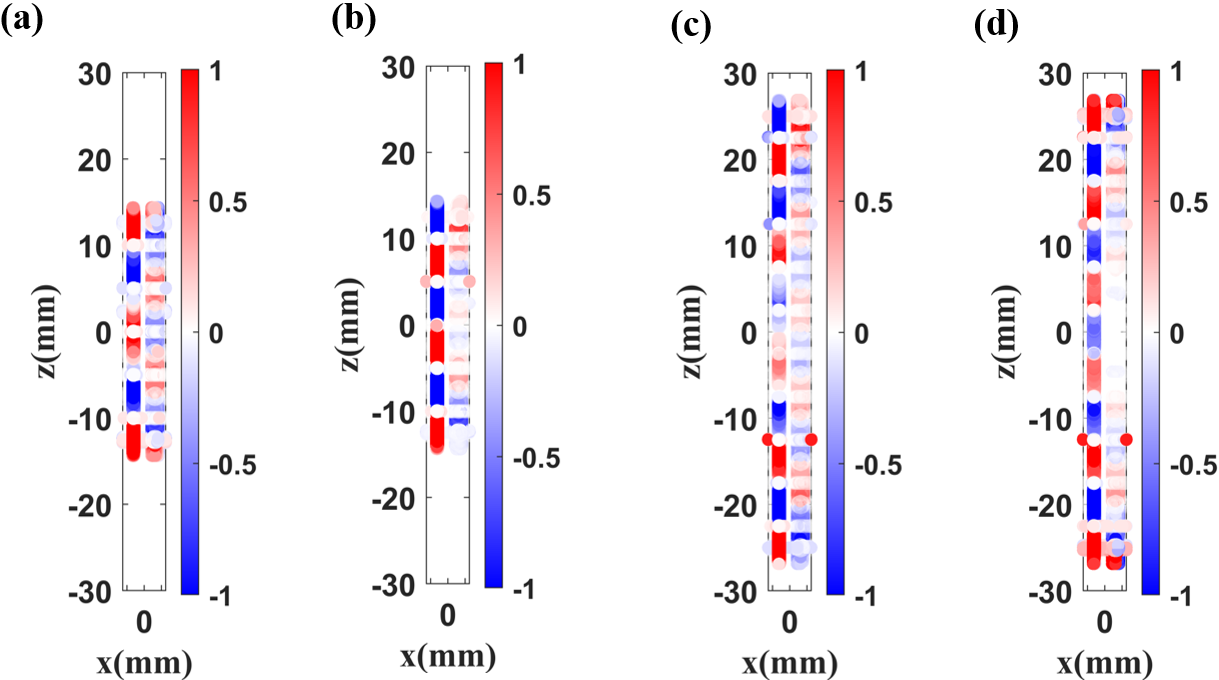}
    \caption{$J_{sz}$ plots of the hybridized modes formed by two TSBICs.
    Even (a) and odd (b) modes for a slab thickness of $N\eq5$;
    odd (c) and even (d) modes for a thickness of $N\eq10$.  }
    \label{fig:TSBICJz}
\end{figure*}

Consider a DNM with a body-centered cubic (BCC) symmetry, which is also called the pcu-c wire-mesh with an offset of $a/2/~ (1,1,1)$ between the two pcu networks.
We here set the radius of each individual net to $\SI{0.08}a$.
A primitive cell, a parallelepiped spanned by the BCC basis vectors, is illustrated in \figref{fig:s10}(a).
The canonical Brillouin Zone, a rhombic dodecahedron, is shown in \figref{fig:s10}(b).
We emphasize the two high symmetry points, invariant under all octahedral ($O_h$) symmetries, $\Gamma\eq(0 , 0 ,0)$ and $H\eq(2\pi/a,0,0)$, connected along the cubic $[100]$ direction.
In the HDP model, we derived a positive slope of  EAW mode, emanating from $\Gamma$ in the long wavelength limit, consistent with the picture in the non-primitive simple cubic Brillouin zone (Section 2 in the main text).
However, the band really emanates at the $H$ point, which is mapped to  the $\Gamma$ point in the simple cubic Brillouin zone.
The correct BCC symmetry classification, with the corresponding bandstructure shown in \figref{fig:s10}(c), thus reveals the left-handed nature of the EAW modes and predicts negative refraction for the $[100]$ inclinated slabs in question.
We note that the left-handed nature of the EAW modes persists in perturbed DNMs of lower simple cubic symmetry with different network offsets or wire radii, even though the bandstructure does not reveal it in these cases.
The underlying reason is rooted in the fact that the dominating Bragg order of the Bloch wave remains outside of the first Brillouin zone close to $(\pm2\pi/a,0,0)$ for these perturbed structures.
We further note that the weak coupling of the quasi-longitudinal EAW modes  to vacuum radiation makes it hard for them to be observed experimentally.

\section{\label{sec:SI8}Topological surface bound states in the continuum }
In Figure 4(b) of the main text, a topological phase transition is observed that leads to topological surface bound states in the continuum (TSBICs), with the corresponding TSBIC bands shown in Figure 4(c) and (d) in the main text.
In the topologically non-trivial region, the distinct parity of the field on the two high symmetry points $X$ (Figure 4(e)) and $\Gamma$ (Figure 4(f))   
leads to a $\pi$ Zak phase in the EAW band.
The corresponding bulk band structure is shown in \figref{fig:s11} (a).

Topological surface bound states emerge in the bulk bandgap due to the nontrivial topological index and a decoupling from vacuum radiation.
In \figref{fig:s11} (b), we show the TSBIC frequencies on the $\Gamma $-point as a function of slab thickness, for integer numbers of unit cells $N$, for which the DNMs keep the geometrically equivalent boundary at the top and bottom surfaces.
For small slab thicknesses, the two surface states on the top and the bottom of the slab are not spatially separated, leading to a frequency splitting of the two \emph{hybridized} modes, which are even/odd with respect to the mirror symmetry at the center of the slab.
Evidently larger thickness of the the DNM slab cause weaker coupling between the two surface modes, hence a reduced energy splitting that converges to zero.
In \figref{fig:TSBICJz}, we show the dominating surface currents $J_{sz}$ of the even/odd hybridized TSBIC states with two different thicknesses of $N\eq5$ and $10$.




\section{\label{sec:SI10} Additional figures}

\begin{figure*}[t]
    \includegraphics[width=.9\textwidth]{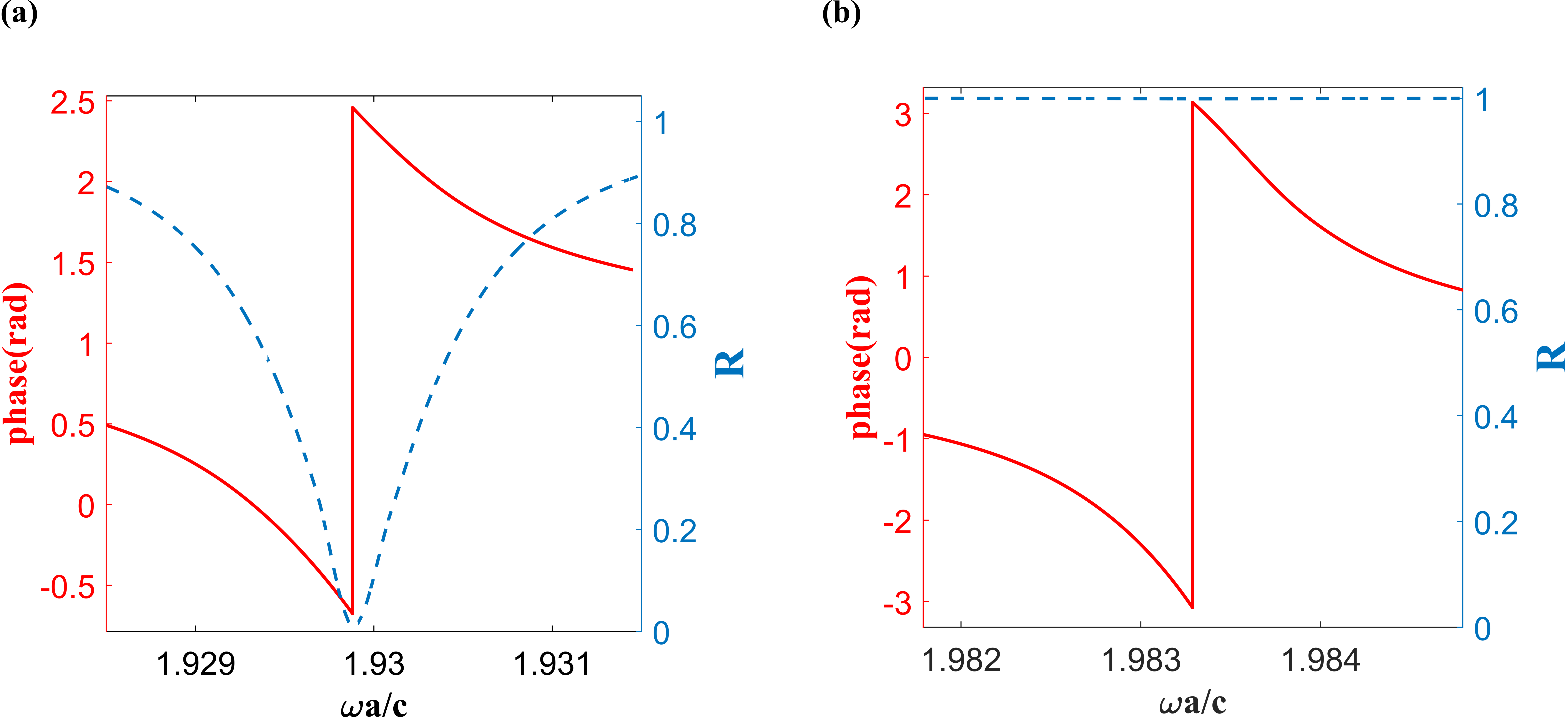}
    \caption{Reflection amplitude (dashed blue) and phase (red) of a symmetric DNM slab with TSBICs on both top and bottom surfaces (a), and a slab with TSBICs only on the reflection side (b).
    (a) Critical coupling with vanishing reflection and a $\pi$ phase jump can be observed at the resonance frequency.
    (b) The reflection is uniformly $1$ across the resonance due to the lack of a transmission channel, while the phase shows a $2\pi$ phase jump indicating an overcoupling regime. }
    \label{fig:Rphase}
\end{figure*}

\begin{figure}[b]
    \includegraphics[width=.6\columnwidth]{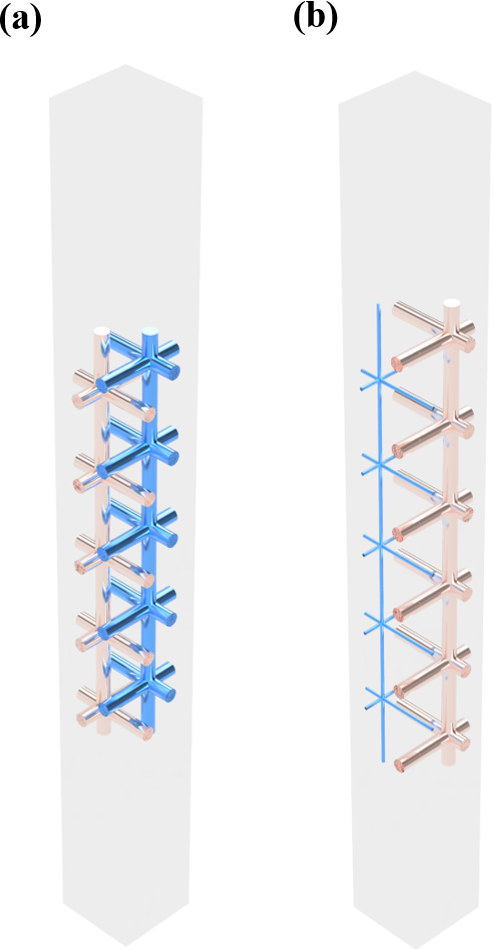}
    \caption{(a) Geometry of DNM slab with $P_1$ parameter; (b) Geometry of DNM slab with different network radii, giving rise to topological surface states. }
    \label{fig:5layergeoms}
\end{figure}

\FloatBarrier

\bibliography{references}

\end{document}